\documentclass{aa}  

\usepackage{graphicx}
\usepackage{txfonts}
\usepackage{amssymb}
\usepackage{amsmath}
\usepackage{float}
\usepackage{placeins}
\usepackage{multicol}
\usepackage{tabularx}
\usepackage{booktabs}
\usepackage[colorlinks=true,citecolor=blue,linkcolor=black]{hyperref}
\usepackage[table,xcdraw]{xcolor}
\DeclareUnicodeCharacter{2061}{}
\begin{document}

   \title{Inferring stellar parameters and their uncertainties from high-resolution spectroscopy using invertible neural networks\thanks{ \href{https://github.com/nils-can/OssicoNN}{https://github.com/nils-can/OssicoNN}}}
   \titlerunning{Stellar parameters with cINN}


   \author{N.~Candebat
          \inst{\ref{arcetri}}
          \and
          G.~G. Sacco\inst{\ref{arcetri}}
          \and
          L. Magrini\inst{\ref{arcetri}}
          \and 
          F. Belfiore \inst{\ref{arcetri}}
          \and
           M. Van der Swaelmen \inst{\ref{arcetri}}
           \and 
           S. Zibetti
           \inst{\ref{arcetri}}
          }

   \institute{INAF -- Arcetri Astrophysical Observatory, Largo E. Fermi 5, I-50125, Florence, Italy\label{arcetri}\\
              \email{nils.candebat@inaf.it}
             }

   \date{today}

 
  \abstract{New spectroscopic surveys will increase the number of astronomical objects in need of characterisation by more than an order of magnitude. Machine learning tools are required to address this data deluge in a fast and accurate fashion. Most machine learning algorithms cannot directly estimate error, making them unsuitable for reliable science.}{We aim to train a supervised deep-learning algorithm tailored for high-resolution observational stellar spectra. This algorithm accurately infers precise estimates while providing coherent estimates of uncertainties by leveraging information from both the neural network and the spectra.}{We trained a conditional invertible neural network (cINN) on observational spectroscopic data obtained from the GIRAFFE spectrograph (HR10 and HR21 setups) within the \textit{Gaia-ESO} survey. A key feature of cINN is its ability to produce the Bayesian posterior distribution of parameters for each spectrum. By analysing this distribution, we inferred stellar parameters and their corresponding uncertainties. We carried out several tests to investigate how parameters are inferred and errors are estimated.}{We achieved an accuracy of 28K in $T_{\text{eff}}$, 0.06 dex in $\log g$, 0.03 dex in $[\text{Fe/H}]$, and between 0.05 dex and 0.17 dex for the other abundances for high-quality spectra. Accuracy remains stable with low signal-to-noise ratio (between 5 and 25) spectra, with an  accuracy of 39K in $T_{\text{eff}}$, 0.08 dex in $\log g$, and 0.05 dex in $[\text{Fe/H}]$. The uncertainties obtained are well within the same order of magnitude. The network accurately reproduces astrophysical relationships both on the scale of the Milky Way and within smaller star clusters. We created a table containing the new parameters generated by our cINN.}{This neural network represents a compelling proposition for future astronomical surveys. These derived uncertainties are coherent and can therefore be reused in future works as Bayesian priors.}

   \keywords{Surveys -- Methods: data analysis -- Stars:abundances -- Stars: fundamental parameters --Techniques: spectroscopic -- 
               }

   \maketitle
%

\section{Introduction}

Most of the recent progress in our understanding of the formation and evolution of the Milky Way is the result of exploration of the wealth of information contained within the catalogues produced by the \textit{Gaia} space mission \citep{Prusti:2016} and provided by a series of ground-based stellar spectroscopic surveys, such as \textit{Gaia-ESO} (GES; \citealt{Randich:2022}), APOGEE \citep{Majewski:2017}, and GALAH \citep{DeSilva:2015}. These surveys have gathered the spectra of hundreds of thousands of stars to derive radial velocities, standard stellar parameters (effective temperatures, surface gravities, and global metallicities), chemical abundances, and other stellar properties (e.g. projected rotational velocities, activity indices). The next generation of multi-object spectrographs and surveys (e.g. SDSS-V \citealt{Kollmeier:2017}; WEAVE \citealt{Shoko:2023}; MOONS \citealt{Gonzalez:2020} and 4MOST \citealt{deJong:2019}) have begun or will soon begin to observe an even greater number of stars than the previous ones ---greater by 
more than an order of magnitude.  

Several new analysis tools based on machine learning techniques have been developed to analyse this huge wealth of data. Some of these new tools, like convolutional neural networks (CNNs;\citealt{Ambrosch:2023}) and AstroNN \citep{Leung:2019}, use artificial neural networks (ANNs) trained with previously analysed and labelled observed spectra to consistently  measure stellar parameters and chemical abundances. We refer to this as the `data-driven' approach. Alternatively, ANNs can be used to interpolate a set of synthetic models in a high-dimensional label space including both stellar parameters and abundances. Other than to infer astrophysical parameters from the spectra, ANNs have been used in stellar spectroscopy for other applications, for example to convert synthetic spectra into observational ones, including the effects of theoretical and instrumental systematic effects \citep{Ting2019, OBriain:2021}.
Spectral analysis codes based on ANNs can process large numbers of spectra much faster than  traditional codes, which fit observed data using a precomputed grid of synthetic spectra derived from atmospheric models (e.g. MATISSE \citealt{Recio2016A&A...585A..93R}; PySME \citealt{Wehrhahn:2023}). Furthermore, they are more efficient at deriving parameters like abundances from low-resolution spectra and combining spectra with photometric and astrometric data \citep{Wang:2023, guiglion:2020, Guiglion:2024}.

While conferring these advantages, this new generation of codes based on ANN for the analysis of stellar spectra are affected by a few limitations. Data-driven codes require training sets made of thousands of spectra that have been independently analysed. Such datasets are not easy to build, especially because they ought to include stars with non-standard properties (e.g. metal-poor or active stars). Also, the results from these codes may be biased because they use astrophysical correlations among parameters during the learning process. For example, the abundance from one element can be inferred from the combination of lines of other elements that are correlated for astrophysical reasons, which can lead to biased results \citep{Ting:2022}. Finally, there is no well-accepted method to calculate errors on the inferred parameters.

Invertible neural networks (INNs), a new ANN architecture proposed by \cite{Ardizzone_INN}, are particularly well suited for estimating physical parameters from observations, because they are designed to provide a posterior probability distribution of the inferred labels. This architecture has already been tested for astrophysical problems ranging from the analysis of the internal structure of planets, to studying the merger history of galaxies, to photometric stellar parameter determination \citep{Haldemann:2023, Eisert:2023, ksoll:2020}. \citealt{Kang:2022, Kang:2023a, Kang:2023b} used a conditional INN to analyse emission lines in H\textsc{ii} regions and stellar parameters from low-mass young stars).

In this study, we developed an INN model for the analysis of medium-resolution (R$\sim$20,000) spectra from GES and derived stellar parameters and some chemical abundances. The paper is organised as follows: In Section~\ref{sec2} we describe the dataset used for training, validation, and testing of the INN model. In Section~\ref{sec3} we present the network architecture, the training process, and the methods used for calculating the errors on the resultant astrophysical parameters. In Section~\ref{sec4} we present the results obtained from applying our code to the GES data, in Section~\ref{sect:astrophysical_validation} we validate our results through some astrophysical tests, and, finally, in Sections ~\ref{sec_discussion} and~\ref{sec_conclusion} we discuss our findings and draw conclusions.

\section{Data}
\label{sec2}

\subsection{The GES GIRAFFE dataset}

In this study, we used data from GES, a large public spectroscopic survey of almost 115,000 stars in the Galactic field and star clusters carried out between 2011 and 2018 with the multi-object spectrograph FLAMES at the VLT \citep{Gilmore:2022, Randich:2022}. All the reduced spectra and the astrophysical parameters derived by the GES consortium are available at the ESO website\footnote{https://eso.org/rm/publicAccess\#/dataReleases}. In particular, the final data release contains spectra observed by the GES consortium or retrieved from the archive with the two FLAMES spectrographs: UVES (R$\sim$47000) and GIRAFFE (R$ \sim$20000). Most of the available setups in both instruments were used during the survey. However, in this work, we used only spectra acquired with the GIRAFFE HR10 (5330-5610 \AA, R=21500) and HR21 (8480-8980 \AA, R=18000) setups, which were used for observing Milky Way field stars as well as clusters and special targets selected as calibrators. The astrophysical parameters used as labels for training, validation, and testing were all retrieved from the last public release (DR 5.1).

The targets catalogue, the data reduction process, the workflow for processing and analysing the data, and the content of the final catalogue of GES are described in \cite{Randich:2022} and \cite{Gilmore:2022}, while the detailed methodologies used for deriving the stellar parameters and chemical abundances of the data release 5.1 are discussed in \cite{Worley:2024} and \cite{Hourihane:2023}.  

\subsection{Dataset pre-processing}

For our analysis, we retrieved the normalised spectra of the HR10 (6060 pixels) and HR21 (10940 pixels) setups  from the GES catalogue and combined them to create spectra with a total of 17000 pixels.  We tested the effect of using a buffer zone to avoid overlapping information between HR10 and HR21, but we found no significant difference in the results. We corrected for the Doppler shift caused by the stellar radial velocity by shifting all the spectra to a common rest frame and applying the flux-conserving method from the \textsc{Astropy} coordinated package \textsc{Specutils}\footnote{see https://specutils.readthedocs.io/en/stable/} for resampling \citep{astropy:2022}. We then filled in any missing values in the normalised flux with the mean flux value.

We then selected only the spectra that have all the stellar parameters used in this work reported in data release 5.1, namely effective temperature (TEFF), surface gravity (LOGG), metallicity (FEH), and elemental abundances of aluminium (AL1), magnesium (MG1), calcium (CA1), nickel (NI1), titanium (TI1), and silicon (SI1). We normalised these parameters to have zero mean and unit variance and fed them into the ANN.
However, to facilitate the visualisation and comparison of our results, we adopted the relative abundance with respect to the solar abundance. The solar abundance values used in this work are taken from \cite{abundance_sun_2007} and listed in Table \ref{tab:sun_element}. Additionally, we may also express the relative abundance with respect to iron, as $\rm [element/Fe]$.

\begin{table}
  \centering
    \caption{Absolute solar abundance and abundance amplitude relative to solar abundances for the stars and elements used in this work.}
  \begin{tabularx}{\columnwidth}{lXcc}
    \toprule \hline
    Element&  & Range [Element/H]\tablefootmark{a} & Solar value \tablefootmark{b} \\
    \midrule
    Aluminium & Al &$[-2.46,1.14]$ dex& 6.37 \\
    Magnesium & Mg &$[-2.46,3.59]$ dex& 7.53 \\
    Calcium & Ca &$[-2.83,1.64]$ dex& 6.31 \\
    Nickel & Ni &$[-3.30,1.37]$ dex& 6.23 \\
    Titanium & Ti &$[-2.90,2.14]$ dex& 4.90 \\
    Silicon & Si &$[-2.21,4.42]$ dex& 7.51 \\
    \bottomrule
  \end{tabularx}
  \tablefoot{
  \tablefoottext{a}{for the stars of the Catalogue dataset.}
  \tablefoottext{b}{Expressed as $\log (N_{el}/N_H)+12$. Solar abundances were computed by \cite{abundance_sun_2007}.}
  }
  \label{tab:sun_element}
\end{table}

Among the ensemble of 114,916 stars encompassed within the GES data release 5.1, a subset of 67,046 stars have been observed with both HR10 and HR21. Among these stars, 11,313 have values for the full set of nine labels used in our analysis. We further impose a cut on the median S/N of 25, empirically identified as offering the best compromise between the quantity and quality of our data. This leaves us with 8391 spectra, of which we use 80\% (6688 spectra) for training and 20\% (1703 spectra) for testing our machine learning models.

At times, we need to compute and analyse the entire dataset. For this, we use the full set of 49,858 stars, which includes all stars observed in HR10 and HR21, with an additional restriction to stars below 7,000K. While this approach might include stars from the training set—which is not ideal—, it becomes necessary due to the limited availability of data. When this occurs, we always compare the results against the GES dataset and look for improvements, as seen in Figs. 10 and 11. Additionally, if other restrictions are applied, they are detailed in the corresponding section. We refer to this dataset as the Catalogue dataset. We have made this catalogue available along with the inferences generated by \texttt{OssicoNN}.

\begin{figure}
    \centering
    \resizebox{\hsize}{!}{\includegraphics{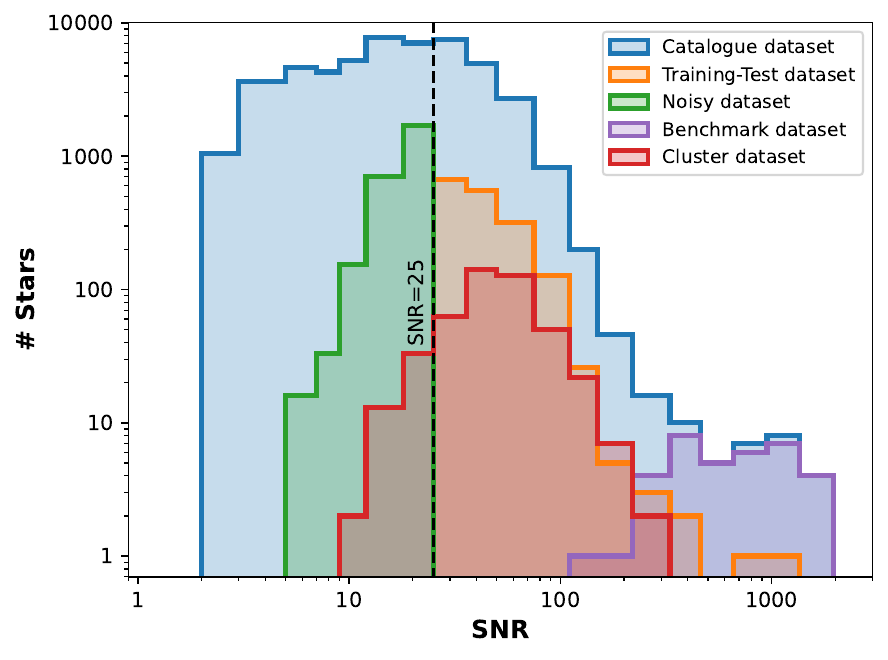}}
    \caption{Distribution of S/N for the various datasets used.}
    \label{fig:snr_dataset}
\end{figure}

To investigate the effects of noise degradation, we constructed a Noisy dataset that consists of 2,592 spectra with a median S/N ranging from 5 to 25. Furthermore, with the aim of validating the results produced by the network, we created two specific datasets. The first, referred to as the Cluster dataset, includes stars belonging to the open clusters NGC 2420, NGC 2243, and Br 32 and globular clusters NGC 1904, NGC 2808, and NGC 362. These clusters include many stars observed with the setups HR10 and HR21, and are well studied and have reliable parameters in the literature \citep{Pancino:2017}. The second dataset, called Benchmark,  contains benchmark stars with well-known parameters in the literature; these stars have been used to calibrate the GES parameters \citep{Worley:2024} and have been observed multiple times, resulting in varying S/N values for the same star. This dataset allows us to validate the reliability of our ANN against other independent methods and observations. The S/N distribution for each dataset is shown in Fig. \ref{fig:snr_dataset}.

\subsection{Data space}
In the Catalogue sample of stars, the effective temperature range is $T_{\rm eff}=3224-6999$ K, the surface gravity log(g) spans from 0.39 to 5.00 dex, and the metallicity $\rm [Fe/H]$ varies from $-$3.47 to 0.49 dex. The chemical abundances also exhibit a wide range of values, which are listed in Table \ref{tab:sun_element}.

\begin{figure*}
\centering
    \includegraphics[width=0.85\linewidth]{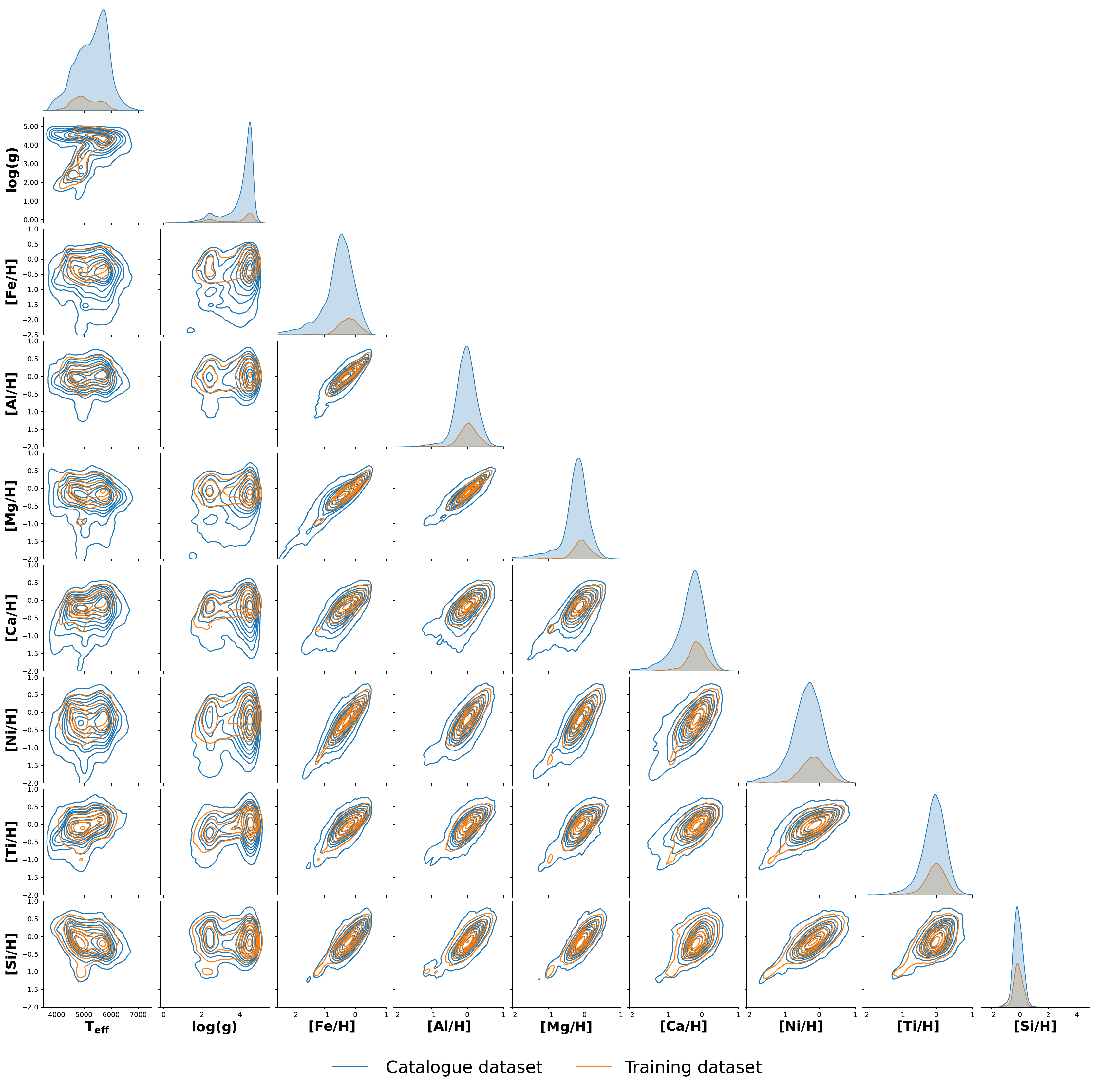}
    \caption{Stellar parameter distribution for the catalogue dataset (in blue) and the training dataset (in orange). The catalogue dataset comprises all stars observed in HR10 and HR21, with an additional restriction to stars below 7,000 K, totalling 49,858 stars. The training dataset contains only the stars from the catalogue dataset used for training (6,688 stars).}
    \label{fig:pairplot}
\end{figure*}

Figure \ref{fig:pairplot} shows the distribution of these parameters and abundances. 
The distributions of effective temperatures and surface gravities are bimodal, while the abundance distributions are close to Gaussian, with negative skewness and a peak below zero. As expected, chemical abundances are positively correlated with each other.
Most of the metal-poor stars included in the full dataset (blue contour) are absent from the dataset used for training and testing (orange contour) because some of their inferred chemical abundances are missing. In the training set (in orange), the temperature range becomes [$3781-6915$]K, the surface gravity spans from 0.51 and 4.84 dex, and the metallicity varies from $-$2.52 to 0.47 dex. Similarly, aluminium ranges from $-$1.60 to  1.06 dex, magnesium from $-$2.40 to  0.74 dex, calcium from $-$2.64 to  0.74 dex, nickel from -2.56 to  1.03 dex, titanium from $-$1.77 to 0.86 dex, and silicon from $-$1.92 to 3.31 dex. The distributions of the Test and Noisy datasets are similar to that of the Training dataset.


\section{Neural network setup}
\label{sec3}

In this section, we describe the architecture of our neural network and how it has been used to derive astrophysical parameters and their errors from the GES spectra. 
In the rest of the paper, we refer to our neural network model as \texttt{OssicoNN}, which takes inspiration from the word ossicones, which refers to the horn-like structures on the head of a  giraffe.

\subsection{Network architecture}

\subsubsection{A nontechnical summary of the architecture}

In this subsection, we summarise the high-level features of the network architecture in nontechnical language. Readers familiar with machine learning terminology may wish to skip this summary and proceed to the following subsections, while readers more interested in the astrophysical applications of our methods may wish to read this summary and then proceed directly to Sect. \ref{sec4}.

The \texttt{OssicoNN} model consists of two linked networks (Fig. \ref{fig:neuralnetwork}): 1) a conditioning network and 2) a conditional INN (cINN). The conditioning network takes as input the observed spectra and extracts high-level features that are then fed into the cINN. Its role is therefore only to process the spectra into a low-dimensional set of variables.

The cINN, on the other hand, is the core of the model and consists of a set of invertible transformations that link the input stellar parameters (labels) to an equal number of hidden variables in a so-called `latent space'. For computational simplicity, variables in the latent space are constrained to be distributed as multidimensional Gaussians. The latent space acts as a reservoir of diversity, allowing us to encode all the information available when moving from spectra to labels, allowing the network to be fully reversible.

The network is trained by feeding it the labels together with the associated spectra, that is, after processing the latter  via the conditioning network. Because of the bijective nature of the cINN, the model also simultaneously learns the inverse transformation, which links spectra and latent space variables to the stellar parameters. When used for inference, \texttt{OssicoNN} exploits this inverse transformation. The inference therefore requires sampling over the Gaussian-distributed variables in the latent space, and, taking the spectra as inputs to the conditioning network, producing an output value for the stellar parameters.

This architecture natively takes parameter degeneracies into account. By repeatedly sampling from the latent space in inference mode, we can recover the posterior distribution function of the stellar parameters generated from a single input spectrum. The resulting (marginalised) uncertainty in the stellar parameters is called internal uncertainty. This uncertainty is linked to the nature of the inverse problem and the ability of the network to learn it, and is not related to the data quality.

We also consider the uncertainty in the model predictions resulting from noise in the data, which we refer to as external uncertainty. This uncertainty is not automatically taken into account by the network, and we explicitly model it by adding realistic noise to the input spectra when running \texttt{OssicoNN} in inverse mode. We model the total uncertainty by summing in quadrature the internal and external terms.

\subsubsection{The invertible neural network approach \label{sect:NN_tech}}
Invertible neural  networks  are part of the family of normalising flow algorithms  \citep{tabak2010}. These algorithms are rooted in the foundational concept of transforming complex probability distributions into more tractable forms, akin to the familiar normal distribution. Normalising flows have found applications in the domain of machine learning, with auto-regressive flows standing out as the foremost examples \citep{kingma2016}. While these normalising flows are theoretically bijective and allow explicit calculation of posterior probabilities, the practical implementation of the reverse transformation proves computationally prohibitive. Addressing this challenge, \citet{Ardizzone_INN} introduced INNs,  endowing the framework with a comprehensive set of properties:

\begin{enumerate}[(i)]
    \item The mapping from inputs to outputs is bijective, meaning that it has an inverse function.
    \item Both the forward and inverse mapping can be computed efficiently.
    \item Both mappings have a tractable Jacobian, which enables the explicit calculation of posterior probabilities.
\end{enumerate}

The fundamental concept underpinning INNs is their ability to alleviate the inherent
degeneracy between the spaces of output labels, denoted as $x$, and observations, denoted as $y$, during the inverse procedure. This is achieved by the introduction of an auxiliary set of variables, referred to as the latent space $z$. Consequently, the neural network transitions from the conventional relationship $f(x)=y$ to a form denoted $f(x) = [y,z]$. The latent space serves as a reservoir, capturing the entirety of the information that becomes lost during the forward process. This augmentation allows a bijective transformation, thus enabling the inverse procedure expressed as $g(y,z)=x$.

To build such INNs, \cite{Ardizzone_INN} combined reversible coupling layers.
Examples of these coupling layers are NICE (Non-linear Independent Components Estimation; \citealt{Dinh:2014}), RealNVP (Real-valued Non-Volume Preserving; \citealt{Dinh:2016}), and GLOW (\citealt{Kingma:2018}; an evolution of RealNVP for image processing). These layers take a vector $\mathbf{u}$ as input, split it into two components $[\mathbf{u_{1}}, \mathbf{u_{2}}]$, and apply the following transformation to produce the output vector $\mathbf{v}$:

\begin{equation}
\begin{split}
\mathbf{v}_{1} & = \mathbf{u}_{1} \odot \exp(s_{1}(\mathbf{u}_{2})) + \exp(t_{1}(\mathbf{u}_{2})), \\
\mathbf{v}_{2} & = \mathbf{u}_{2} \odot \exp(s_{2}(\mathbf{v}_{1})) + \exp(t_{2}(\mathbf{v}_{1})),
\end{split}
\end{equation}
where $\odot$ denotes element-wise multiplication and + is vector addition. 

The block's reversibility allows a straightforward conversion from the latent space vector $\mathbf{v}=[\mathbf{v_{1}},\mathbf{v_{2}}{}]$ back to the original vector $\mathbf{u}$ through the following transformation:
\begin{equation}
\begin{split}
\mathbf{u}_{2} & = (\mathbf{v}_{2} - t_{2}(\mathbf{v}_{1}) \odot \exp(-s_{2} (\mathbf{v}_{1})), \\
\mathbf{u}_{1} & = (\mathbf{v}_{1} - t_{1}(\mathbf{u}_{2}) \odot \exp(-s_{1} (\mathbf{u}_{2})).
\end{split}
\end{equation}

The reversible blocks in our model use the internal functions $s_{i}$ and $t_{i}$ ---which are neural networks of our design (typically composed of a linear layer  and a rectified linear unit (ReLU) activation for simplicity)--- to map the parameter space to the latent space (and back). The benefit of these blocks is that the $s_{i}$ and $t_{i}$ networks always operate in the forward direction,  regardless of whether we are performing the forward or the inverse transformation. Not only does this make it possible to calculate both forward and inverse transformations, but as the hyperparameters of the linear layers are the same in both forward and inverse modes, both transformations are learned simultaneously.

Subsequent to each reversible block, a deterministic permutation is systematically applied to the vector $\mathbf{u}$ (or $\mathbf{v}$ in the inverse mode), introducing a stochastic yet fixed rearrangement of vector elements. This permutation serves the dual purpose of enhancing interactions among vector elements, thereby enriching the expressiveness of the model, while maintaining a deterministic footing to ensure reproducibility and stability in the learning process. Ultimately, these computational blocks have the advantageous property of allowing for an easy computation of the determinant of the Jacobian matrix of the transformation.

\subsubsection{Conditional invertible neural networks}
Invertible neural  networks faced limitations due to rigid dimensionality requirements between $x$, $y$, and $z$. This was particularly challenging when $x$ and $y$ had significantly different dimensionality, like in the case where a few physical parameters are inferred from spectra with a large number of spectral channels. This limitation is alleviated in conditional invertible neural networks (cINNs, \citealt{Ardizonne:cINN}). While the fundamental idea of building a transformation between spaces remains, the innovation of cINNs  lies in constraining this transformation between hidden parameters $x$ and the latent space $z$ using a condition ($c$) derived from observations $y$.

During training, this transformation is expressed as $z = f(x; c=y)$, and the subsequent inverse transformation recovers hidden parameters as $x= g(y;c=y)$. This refined architecture enables flexibility by decoupling the dimensions of $x$ and $y$, allowing work with a latent space $z$ aligned with the dimensions of $x$.

Practically, this new condition $\mathbf{c}$ is seamlessly integrated into the $s_{i}$ and $t_{i}$ neural networks in the reversible blocks. This is achieved by merely combining the condition vector c with the input vector ($\mathbf{u}$ during forward transformation, $\mathbf{v}$ in inverse mode) before entering the neural network. Therefore, the notation $s_{i}(\mathbf{u}_{j})$ is replaced by the notation $s_{i}(\mathbf{u_{j}}, \mathbf{c=y})$ in the cINN framework.

\subsection{Implementation of conditional
invertible neural networks }
In scenarios involving observations with high-dimensional data, such as spectra and images, including  them in reversible blocks becomes computationally expensive and inefficient. This is primarily due to the substantial number of pixels, many of which might not contribute useful information. Therefore, we introduce a conditioning neural network $h$, which is a function that maps the observations to a lower-dimensional representation. This representation captures the salient aspects of the data that are related to the hidden parameters $x$.

The whole cINN is represented in Fig. \ref{fig:neuralnetwork}. The architecture consists of two distinct parts. The first is the conditional network, which is composed of different reversible blocks. During training, information travels in the forward direction through the conditional network along the blue arrows from star parameter labels to latent space. During inference, the conditional network operates in reverse mode and information travels along the red arrows. The second part of the architecture is the conditioning network. Comprising layers linked by green arrows, this network processes the input spectra and extracts relevant features. It enables efficient computation by injecting information at various stages within the coupling blocks of the conditional network.

We used \textsc{Freia}\footnote{\href{https://github.com/vislearn/FrEIA}{https://github.com/vislearn/Freia}} \citep{freia}, a library that facilitates the design of INNs (and cINNs) based on the \textsc{Pytorch} library \citep{torch}, to construct the entire network. The GLOW blocks were implemented following the logic explained in the documentation, and we adopted the layers and logic of Pytorch.

Hyperparameters are the user-defined settings that control the learning process of a model, such as the number of hidden layers, the learning rate, and the regularisation strength. The choice of hyperparameters can have a significant impact on the performance and accuracy of a model, but it is often a time-consuming and tedious task to manually tune them. To overcome this challenge, we used \textsc{Optuna} \citep{optuna_2019}, a software framework for automated hyperparameter optimisation. \textsc{Optuna} provides a flexible and easy-to-use interface for defining and exploring the hyperparameter space and finds the best hyperparameters. In particular, we employed the Tree Structured Parzen Estimator (TPE) sampler ---which is based on the Bayesian method--- to systematically explore the parameter space with 100 distinct models.

We tasked \textsc{Optuna} with the objective of minimising the mean squared error (MSE) between GES values and inferred values. To attain this goal, we  maintained the structure  of the conditioning neural network  while adjusting the parameters of the layers:
\begin{itemize}
    \renewcommand{\labelitemi}{$\bullet$}
    \item Convolutional layer :
    \begin{itemize}
        \item Convolution\_out [$2^{i}, i \in [2,8]$] ,
        \item Convolution\_kernel [$2^{i}, i \in [2,8]$] ,
        \item Convolution\_stride [$1,2$]. 
    \end{itemize}

    \item Maxpool :
    \begin{itemize}
        \item maxpool [$1,2,3,4$].
    \end{itemize}

    \item Dense :
    \begin{itemize}
        \item Dense\_out [$2^{i}, i \in [2,8]$] .
    \end{itemize}
\end{itemize}

The selection of hyperparameters is thus grounded in either optimisation or, if explicitly indicated, the recommendations provided by the creators of the cINN \citep{Ardizonne:cINN}.

\subsubsection{Invertible network}
The hidden data space and the latent space are bridged by the INN, which consists of four GLOW blocks. Each GLOW block contains two identical neural networks, $s$ and $t$, which are composed of two dense layers with 512 neurons each with ReLU activations. The outputs of the cINNs are concatenated to these dense layers. Prior to each GLOW block, a random permutation block is inserted as previously mentioned, and  subsequent to each block, a normalisation block is positioned. This design adheres to the guidelines proposed by the original authors of the cINN.

\subsubsection{Conditioning network}
The conditioning network follows a standard CNN setup, with 1D convolutional layers using ReLU activation. After the first convolutional layer, we also add a subsequent max-pooling to reduce output size. This structure repeats three times, followed by flattening and three linear layers with ReLU activation and dropout regularisation. The output then goes through a final linear layer before joining the last GLOW block.

\begin{figure*}
    \centering
    \resizebox{\linewidth}{!}{\includegraphics[angle=0]{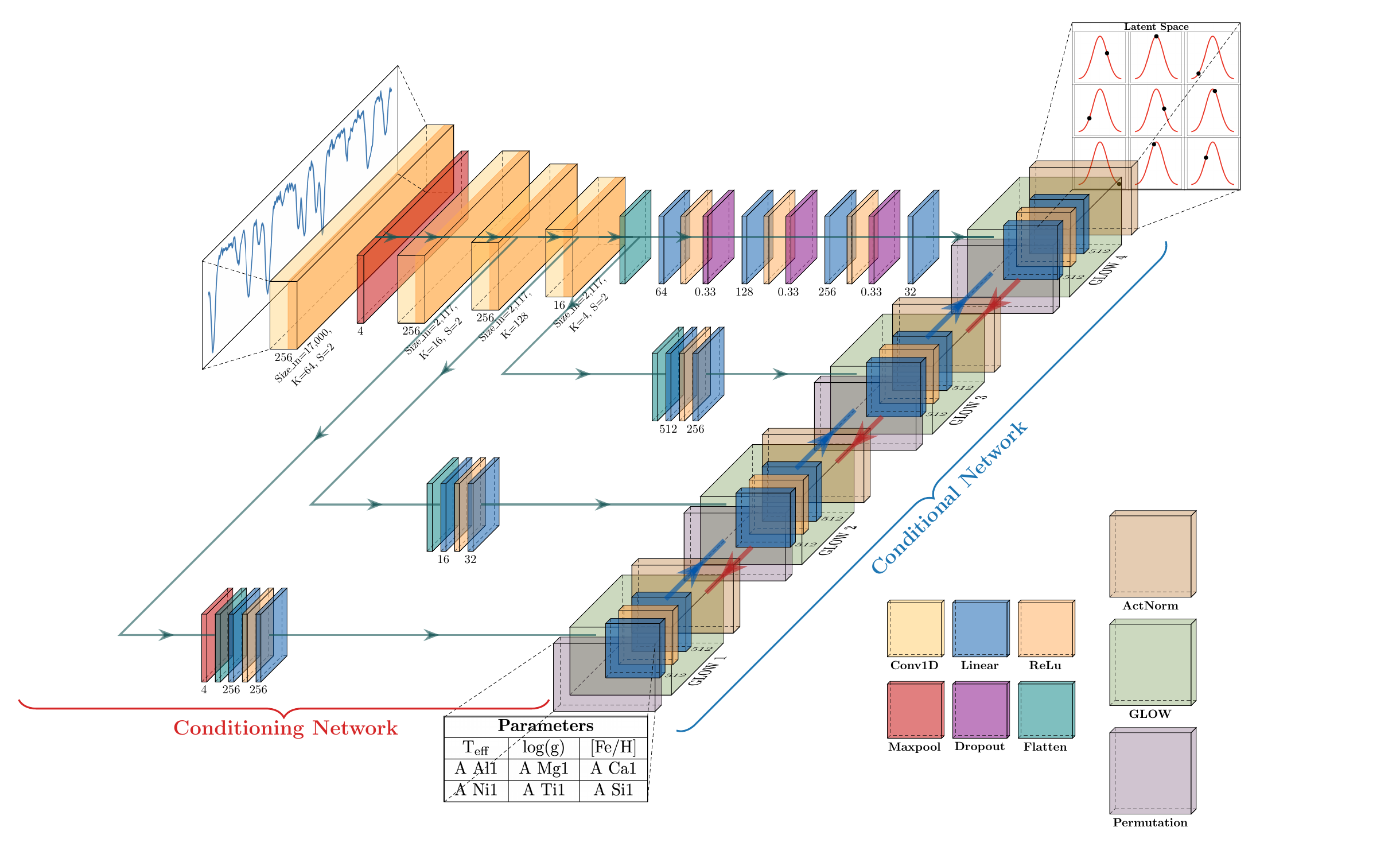}}
    \caption{Schematic representation of \texttt{OssicoNN}, the neural network for the inference of stellar parameters. The primary objective of the neural network is to encapsulate a transformative mapping between the hidden stellar parameters and an abstract latent space composed of an equivalent number of Gaussian distributions. This is done by the conditional network (in blue). The mapping is conditioned by the spectra first processed by the conditioning network (in red) that extracts the important features from the spectra. Each layer is shown as a box in a different colour, reflecting its nature, with key hyperparameters listed beside it.}
    \label{fig:neuralnetwork}
\end{figure*}

Moreover, outputs from the second, third, and fourth convolutional layers are duplicated, flattened, and processed through two linear layers with ReLU activation (except for the first  one, which also adds a max-pooling layer). These processed outputs serve as conditioning inputs for the first, second, and third GLOW blocks, respectively. The feature numbers at each convolutional layer stage are 256, 32, 256, and 32, respectively. These features play a key role in conveying relevant spectral information to the conditional network, influencing how data space is mapped to latent space. The complete set of hyperparameters of the different layers is shown in Fig. \ref{fig:neuralnetwork}.

\subsubsection{Training process}
\cite{Ardizonne:cINN} suggest that the cINNs be trained with a maximum likelihood loss based on input vector parameters, latent space parameters, and model parameters $\theta$:

\begin{equation}
    \mathcal{L} = \mathbb{E}_{i}\left[\frac{||f(x_i;\theta, y_i)||^{2}}{2} - \log \left( \left| \det\left(\frac{\partial f}{\partial x}\right)\right|\right) \right] + \frac{1}{2\sigma^2}||\theta||^{2}.
    \label{loss_text}
\end{equation}

The final loss function therefore consists of two components: the first one is the maximum likelihood objective, and the second one is an L2 regularisation term. In practice, we only implement the maximum likelihood objective, and the regularisation term is added separately.

To facilitate the training process, we adopt the other techniques proposed by the cINN authors. First, we apply soft clamping of the hyperparameters by constraining the gradient norm below 10. We additionally use the `clamp' parameter within the GLOW layer specifications. This parameter restricts the exponential term in equations (2) and (3) to $exp(\pm s_{clamp})$. Following the\cite{Ardizonne:cINN} recommendation we chose $s_{clamp}=1.9$.
These decisions are driven by the consideration that the exponential function is employed in coupling layers, which poses a risk of amplifying the values of elements in \textbf{u} (or \textbf{v}), potentially resulting in divergence. Second, we use soft channel permutations (i.e. random orthogonal matrices) to allow the exchange of information between $\mathbf{u_1}$ and $\mathbf{u_2}$ within the blocks. Third, we use normal initialisation with a scaling factor of 0.01, which improves the initial stage of training. Moreover, when training, we introduce Gaussian noise with an amplitude of 0.05 to the hidden parameters in the training set to improve performance and general stability.

We train these hyperparameters using the loss function in Eq. \ref{loss_text}, and we additionally compute the mean squared error (MSE) as another metric with which to monitor the training progress. We compute these two metrics for both the training and the validation dataset. During training, spectra are fed to the neural network in batches of 32 elements. We train for 60 epochs with a learning rate of lr=$10^{-3}$, which is reduced by a factor of ten at epoch 50. Furthermore, we apply a weight decay of 1e-5 to the model parameters as a form of L2 regularisation, which helps to prevent overfitting and improve generalisation performance.

\begin{figure}
    \centering
    \includegraphics[width=\hsize]{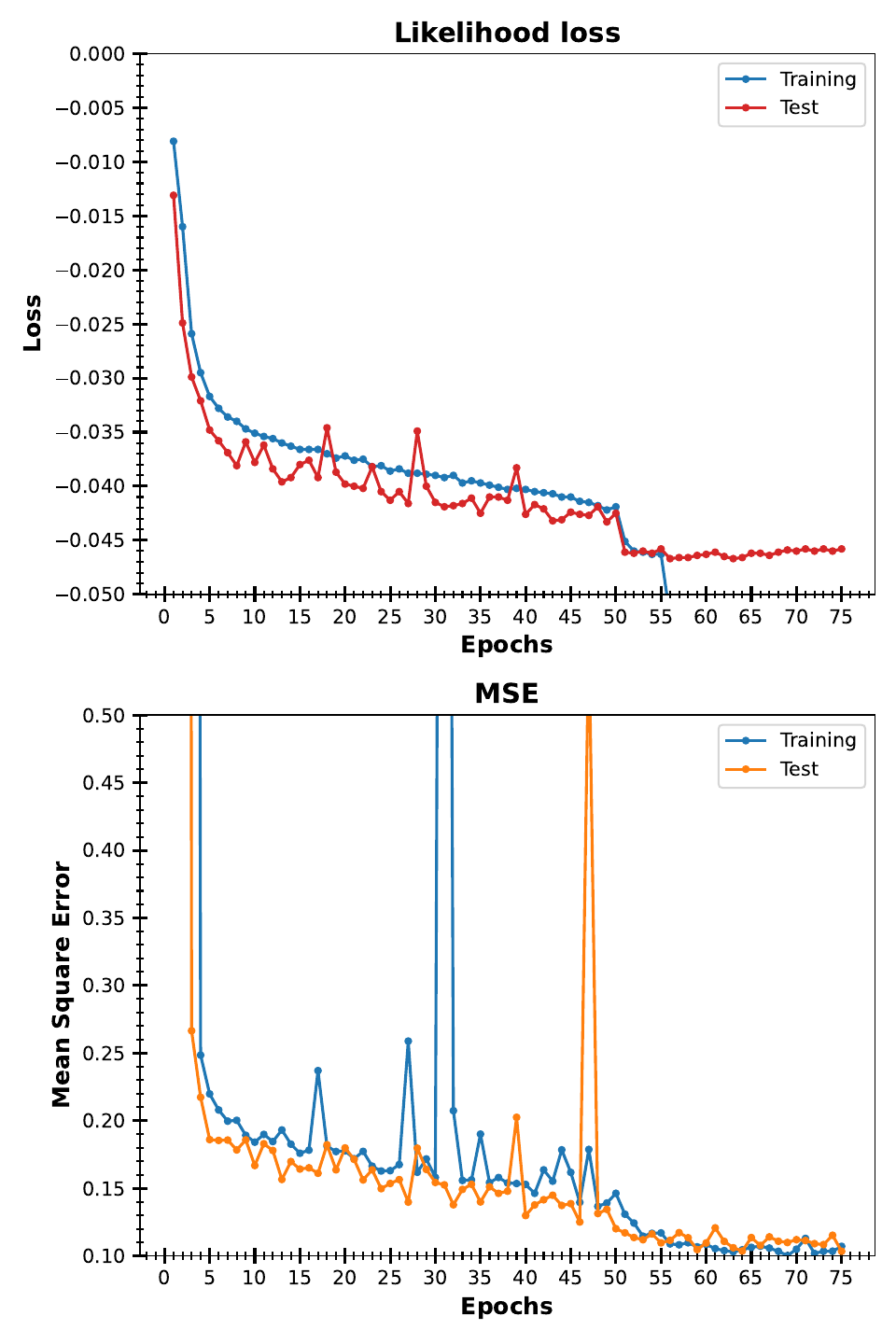}
    \caption{Evolution of the loss function and the MSE during training. The learning rate was reduced from $10^3$ to $10^4$ in period 50, leading to a drop in overall losses.}
    \label{fig:loss_training}
\end{figure}

The training of our neural network model with a NVIDIA RTX4090 takes approximately 25 minutes for 60 epochs. The evolution of these different metrics during training is shown in Fig. \ref{fig:loss_training}. The MSE exhibits instability, particularly within the initial 50 epochs when the learning rate is set to lr$=$0.001. This behaviour is not unexpected, considering the inherent dissimilarity between the MSE error metric and the likelihood function. While the MSE error term penalises larger deviations due to its quadratic nature, the likelihood function focuses on points that deviate from the underlying data probability distribution.

Training neural networks inherently involves randomness due to the different initial conditions and the random sequencing of training data exposure.To ensure the selection of a well-trained network, we trained five versions of the model and opted for the one that obtains the least loss.

\subsection{Uncertainties \label{sec_uncertainty_theory}}

Supervised machine learning algorithms represent a distinct paradigm compared to `classical' algorithms. Classical algorithms derive results from predefined rules and observations, whereas in supervised machine learning algorithms, the results and observations are employed to deduce rules (and potentially generate new results). The nature and origin of uncertainties are therefore very different. Consequently, various methods have been suggested to address uncertainties within this alternative paradigm \citep{hullermeier:2019}. The theoretical approach to calculating these uncertainties relies on quantifying the disparity between an ideal transformation and one derived from our neural network and data.

This gap leads to the emergence of various types of uncertainty \citep{Levasseur:2017}. Here, we primarily consider epistemic and aleatoric uncertainties. Epistemic uncertainties arise from errors inherent to the model, specifically due to the lack of knowledge about the perfect model. In an ideal scenario, these uncertainties can be reduced with additional information. Such uncertainties include, among other factors, model uncertainty, indicating that the neural network, constrained by the chosen architecture, may not precisely emulate the perfect transformation. For instance, a neural network composed solely of dense layers without activations might only model linear transformations, resulting in a mismatch with the perfect transformation when the mapping between observation space $Y$ and output space $X$ is not itself linear. Epistemic uncertainties also include approximation uncertainty, which arises due to limitations in the learning process, such as halting after a finite number of iterations or  populations inadequately represented in the dataset with respect to $X \times Y$.

Furthermore, aleatoric uncertainties emerge from the inherent variability caused by randomness. In the case of astrophysical spectra and INN, this will include factors such as photon noise, potential data corruption, calibration uncertainties, observational artefacts, potential errors in the hidden parameters, and the non-deterministic relation between observations and parameters. These uncertainties are similar in that improvements by adding new information ---whether through alterations to architecture and training methods or by increasing the dataset--- are ineffective in reducing them.

Addressing all uncertainties in our analysis is quite complex. We address this challenge by computing two types of uncertainties. The first, referred to here as internal uncertainty, is a combination of some epistemic and aleatoric uncertainties associated with the neural network. The second, which we call external uncertainty, accounts for uncertainties in data quality.

\subsubsection{Internal uncertainties in cINN}

Internal uncertainty encompasses the inherently non-deterministic relationship between spectrum and parameters. Even with complete information about spectrum, there remains uncertainty in the predicted parameters, which is considered as aleatoric uncertainty here. This effect is further amplified as data are observational (and therefore inherently noisy, even after curation) and parameters are derived from a classical pipeline that introduces its own biases. Additionally, internal uncertainty considers part of the approximation uncertainty created by the quality and density of the training data.

However, internal uncertainties do not encompass all components of epistemic uncertainty. For example, we do not account for the full range of model uncertainties or certain aspects of approximation uncertainties. This is because we maintain specific assumptions about the model, such as keeping the weights fixed and not measuring the impact of changes induced by adjusting the model's weights.


Computing these internal errors is achieved by determining the posterior distribution of the parameters according to the observations $p(x|y)$. To do this, we use the fact that cINN models a transformation between the distribution of the parameters and a normal distribution $p(z)=N(z,0,I)$. We can therefore obtain the posterior distribution by sampling the normal distribution in $N_{\mathrm{internal}}$ samples and performing the inverse transformation $x=g(z,c=y)$. This exploration of the full latent space through $N_{\mathrm{internal}}$ samples allows us to capture the complete diversity of possible solutions for a given spectrum, thereby reconstructing the full Bayesian posterior for all parameters.

Marginalising along a small number (1 or 2) of parameters allows us to effectively explore the full multivariate posterior. For the sake of visualisation, we studied the pairwise distribution of all parameters. The distributions for three spectra Figs. \ref{fig:distribution_latent_space_20}, \ref{fig:distribution_latent_space_250}, \ref{fig:distribution_latent_space_750} are provided in appendix.

The posterior distribution of each parameter is unimodal and approximately Gaussian, indicating that the problem we are addressing does not exhibit strong degeneracies. We studied the pairwise parameter distributions, finding unimodal, bivariate, Gaussian-like distributions, with pairs of parameters being weakly correlated or not at all. We calculated the Pearson coefficients for all stars in the test set (see Fig. \ref{fig:distri_pearson_resume} and Fig. \ref{fig:distribution_pearson_coeff} in appendix for full coefficients) and determined that the most highly correlated parameters are temperature and surface gravity (weak linear correlation, median $R=0.3$) and temperature and metallicity (very weak correlation, median $R=0.18$). The median Pearson coefficients for other parameter pairs are below 0.14.

\begin{figure}
    \centering
    \includegraphics[width=\hsize]{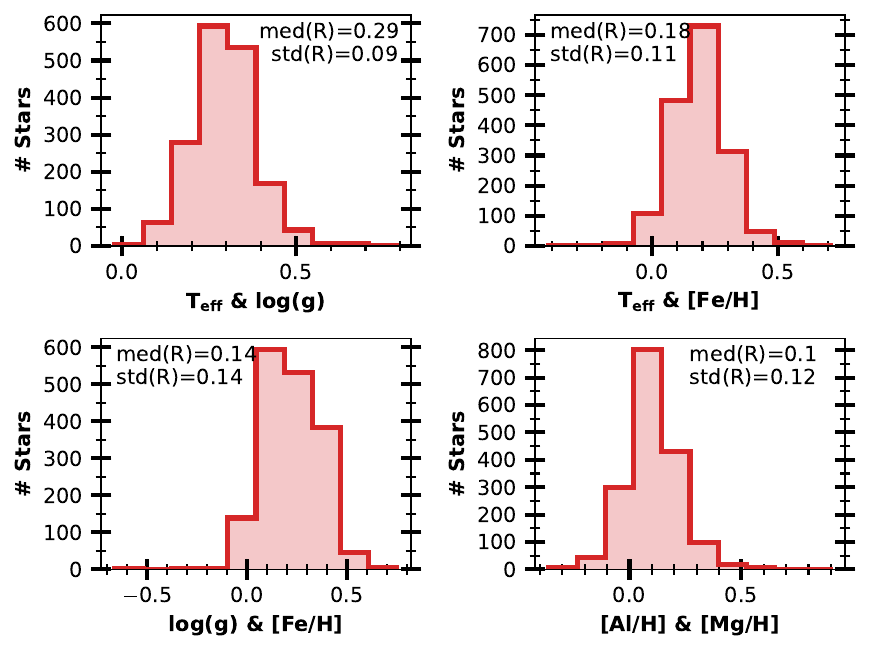}
    \caption{Distribution of the Pearson coefficients between pairs of parameters for test set stars derived from sampling latent space to obtain the posterior distribution for one spectrum. Temperature, metallicity, and gravity exhibit the strongest correlations (albeit weak), whereas abundances, such as [Al/H] and [Mg/H], show very poor correlations. Additional Pearson coefficient distributions are provided in appendix Fig. \ref{fig:distribution_pearson_coeff}.}
    \label{fig:distri_pearson_resume}
\end{figure}

\subsubsection{External uncertainties in cINN}

External uncertainties encompass all the sources of error in the data. In this study, we account for observational errors that affect the data quality, such as Poisson noise from photon counts on the CCD, cosmic-ray hits, atmospheric effects, and so on. However, we do not consider systematic errors that are intrinsic to the instrument and affect all spectra in a similar way. We also assume that the stellar parameters assigned to each spectrum by GES are accurate and reliable, without any bias or misclassification.

We estimated external uncertainties from the GES data products. The GES data reduction pipeline produces the inverse of the variance(IVAR)  for each pixel. To obtain the standard deviation of the noise, we take the square root of the inverse of IVAR multiplied by the ratio between the normalised FLUX\_NORM and the non-normalised flux (FLUX):
\begin{equation}
    \sigma_{\mathrm{pixel}} = \sqrt{\frac{FLUX}{FLUX\_NORM} \times \frac{1}{IVAR\_FLUX}}
    \label{sigma_pix}
.\end{equation}
This ensures that we estimate the noise level of the normalised spectrum.

To account for external uncertainty in our machine learning models, we generated $N_{\mathrm{external}}$ noisy spectra for each observed spectrum by adding Gaussian noise with a mean of zero and a standard deviation $\sigma_{\mathrm{pixel}}$ (Eq. \ref{sigma_pix}) to each pixel $N(0,\sigma_{\mathrm{pixel}}^2)$: 
\begin{equation}
    y_{noisy} = y_{GES}+ \mathrm{random}(\mathcal{N}(0, \sigma_{pixel}^2))
.\end{equation}
We assume that the pixels are independent and we neglect the covariance between the pixels.
This procedure creates a distribution of spectra that represents the possible variations due to noise. This method is similar to that applied by \cite{Kang:2022} to the HII region, although our treatment of uncertainties is different, as we separate the internal and external uncertainties.

\subsubsection{Total uncertainties}
\label{sec_total_uncertainties}
To estimate the total uncertainty, we combined the two sources of error: internal and external. We first generated $N_{\mathrm{external}}$ spectra with added noise, simulating the observational uncertainty. For each noisy spectrum, we drew $N_{\mathrm{internal}}$ samples from the latent space, representing the diversity of the neural network. We then applied \texttt{OssicoNN} in inverse mode to infer the stellar parameters for each spectrum and latent sample. This gives us the posterior distribution of all the parameters  for each $N_{external}$ spectrum in the form tensor of shape $(N_{\mathrm{stars}}, N_{\mathrm{external}},N_{\mathrm{internal}}, N_{\mathrm{parameters}})$. 

To find the total error for each parameter, we characterised the distribution differences.
We first looked at external uncertainty, that is, how the distribution changes between $N_{\mathrm{external}}$ iterations of the same spectra. The most reliable method entails using kernel density estimation (KDE) and identifying the peak in the density distribution for each of the $N_{\mathrm{external}}$ spectra. To do this, we used the \textsc{Scikit-learn} kernel density function with a Gaussian kernel and a bandwidth determined by the Silverman rule. We recorded the peak of the density for each spectrum. The external error is directly derived by computing the standard deviation on the subset composed of values linked with peak density:

\begin{equation}
    \epsilon_{external} = \sigma(\mathrm{argmax}(\mathrm{KDE}(x_{N_{internal}})))_{N_{external}}
.\end{equation}

The internal uncertainty is derived by measuring the amplitude of the posterior distribution when sampling the latent space. We measured this global amplitude by calculating the interval $u_{68}/2$ of each posterior distribution, where $u_{68}$ denotes the interval between percentile 16 and percentile 86 of the distribution. The final value is determined by calculating the median along the $N_{\mathrm{external}}$ values: 

\begin{equation}
    \epsilon_{\mathrm{internal}} = \mathrm{median}(u_{68}(x_{N_{\mathrm{internal}}})/2)_{N_{\mathrm{external}}}
.\end{equation}

The overall uncertainty is computed by quadratically summing these two distinct uncertainties:

\begin{equation}
    \epsilon_{\mathrm{total}} = \sqrt{\epsilon_{\mathrm{internal}}^2 + \epsilon_{\mathrm{external}}^2 }
.\end{equation}

\subsection{Inference of the parameters}
The algorithm offers two modes of inference that differ in their speed and accuracy. In the Fast mode, the algorithm randomly selects a single sample from the latent space for each dimension according to a Gaussian probability. This sample, together with the spectrum to be analysed, is fed to the neural network in inverse mode, which then infers all the parameters of the associated spectrum. On the other hand, the Precision mode allows more precise parameter estimation and measurement of the uncertainties discussed earlier. To estimate parameter values, both IVAR information and sampling of the latent space are employed to obtain the same tensor as previously described in Section \ref{sec_total_uncertainties}. The final task involves converting the parameter distribution into a singular value. The most reliable method again entails using KDE and identifying the peak of the posterior distribution. Although this step is resource-intensive, it is indispensable when dealing with multimodal distributions resulting from certain physical degeneracies. 
However, in the case of the stellar parameters under investigation, this complexity is absent. Therefore, to validate our estimation, we could also have taken the median of the distribution to save time. Finally, the median is calculated over the $N_{\mathrm{external}}$ samples.

\begin{equation}
    x_{OssicoNN} = \mathrm{median}(\mathrm{argmax}(\mathrm{KDE}(x_{N_{\mathrm{internal}}})))_{N_{\mathrm{external}}}.
\end{equation}

\texttt{OssicoNN} achieves a rate of 5000 iterations per second for parameter determination of a ${spectrum-random\_latent\_space}$ pair. This efficiency enables rapid predictions during real-time analysis.  In Fast mode, analysing a catalogue of about 50,000 stars takes roughly 10 seconds. However, switching to Precision mode incurs a significant increase in computational overheads. Given that we utilise $N_{\mathrm{external}}$ spectra and $N_{\mathrm{internal}}$ samples from the latent space for each spectra,  the computation time scales with the product of these two parameters. For instance, with $N_{\mathrm{external}}=20$ and $N_{\mathrm{internal}}=250$, processing the same catalogue would require approximately 14 hours.

\begin{table*}
  \centering
  \addtolength{\tabcolsep}{-3pt}
    \caption{Residual statistics between \texttt{OssicoNN} and GES for various datasets and two modes: Fast and Precision.}
  \begin{tabular}{c c c c c c c c c c c c c c c c c}
    \hline
    \hline
      & & \multicolumn{5}{c}{$\mathbf{T_{eff}}$} & \multicolumn{5}{c}{\textbf{log(g)}} & \multicolumn{5}{c}{\textbf{[Fe/H]}}\\
    \hline
       Dataset&mode& med & mean & std & p15 & p85 & med & mean & std & p15 & p85 & med & mean & std & p15 & p85\\
    \hline
    Test&F& +1 & +1 & 34 & -27 & +27 & +0.00 & +0.00 & 0.07 & -0.06 & +0.06 & 
    +0.00 & +0.00& 0.03 &  -0.03 & +0.03\\
    Test&P& +1 & +1 & 28 & -19 & +23 & +0.00 & +0.00 & 0.06 & -0.04 & +0.04 & 
    +0.00 & +0.00& 0.03 &  -0.02 & +0.02\\
    Noisy&F& +4 & +6 & 43 & -30 & +43 & +0.00 & +0.00 & 0.08 & -0.06 &+0.07 & 
    +0.00 &+0.00& 0.05 &  -0.05 & +0.03\\
    Noisy&P& +1 & +10& 39 & -32 & +32 &
    +0.00 & +0.00 & 0.08 & -0.06 & +0.07 & 
    +0.00 & +0.00& 0.05 &  -0.03 & +0.04\\
    \hline
    \hline
      & & \multicolumn{5}{c}{\textbf{[Al/H]}} & \multicolumn{5}{c}{\textbf{[Mg/H]}} & \multicolumn{5}{c}{\textbf{[Ca/H]}}\\
    \hline
      Dataset&mode& med & mean & std & p15 & p85 & med & mean & std & p15 & p85 & med & mean & std & p15 & p85\\
    \hline
    Test&F& -0.01 & -0.01 & 0.06 & -0.07 & +0.05 &
    -0.01 & +0.00 & 0.05 & -0.06 & +0.04 & 
    +0.00 & +0.00& 0.18 &  -0.16 & +0.16\\
    Test&P& +0.00 & +0.00 & 0.05 & -0.05 & +0.04 &
    +0.00 & +0.00 & 0.04 & -0.04 & +0.04 & 
    +0.01 & +0.01 & 0.13 & -0.11 & +0.13\\
    Noisy&F& +0.00 & -0.01 & 0.10 & -0.08 & +0.07 &
    +0.00 & -0.01 & 0.07 & -0.07 & +0.06 & 
    +0.01 & +0.01& 0.24 &  -0.20 & +0.23\\
    Noisy&P& -0.01 & +0.00 & 0.09 & -0.06 & +0.06 &
    -0.01 & -0.01 & 0.06 & -0.06 & +0.05 & 
    +0.02 & +0.01& 0.20 &  -0.16 & +0.20\\
    \hline
    \hline
      & & \multicolumn{5}{c}{\textbf{[Ni/H]}} & \multicolumn{5}{c}{\textbf{[Ti/H]}} & \multicolumn{5}{c}{\textbf{[Si/H]}}\\
    \hline
      Dataset& mode & med & mean & std & p15 & p85 & med & mean & std & p15 & p85 & med & mean & std & p15 & p85\\
    \hline
    Test&F& -0.01 & +0.01 & 0.18 & -0.18 & +0.17 &
    +0.00 & +0.00 & 0.18 & -0.15 & +0.15 & 
    +0.00 & +0.00& 0.22 &  -0.11 & +0.12\\
    Test&P& +0.01 & +0.01 & 0.14 & -0.12 & +0.13 &
    +0.01 & +0.01 & 0.13 & -0.11 & +0.12 & 
    +0.00 & +0.00& 0.15 &  -0.08 & +0.10\\
    Noisy&F& -0.01 & +0.00 & 0.28 & -0.26 & +0.28 &
    -0.01 & -0.01 & 0.25 & -0.24 & +0.21 & 
    +0.00 & -0.01& 0.25 &  -0.17 & +0.15\\
    Noisy&P& +0.01 & +0.01 & 0.24 & -0.21 & +0.24 & -0.01 & -0.01 & 0.21 & -0.19 & +0.17 & 
    -0.01 & -0.01& 0.19 &  -0.13 & +0.13\\
    \hline
  \end{tabular}
  \label{tab:param_ossiconn}
\end{table*}

   \begin{figure*}
    \centering
   \resizebox{\linewidth}{!}
            {\includegraphics{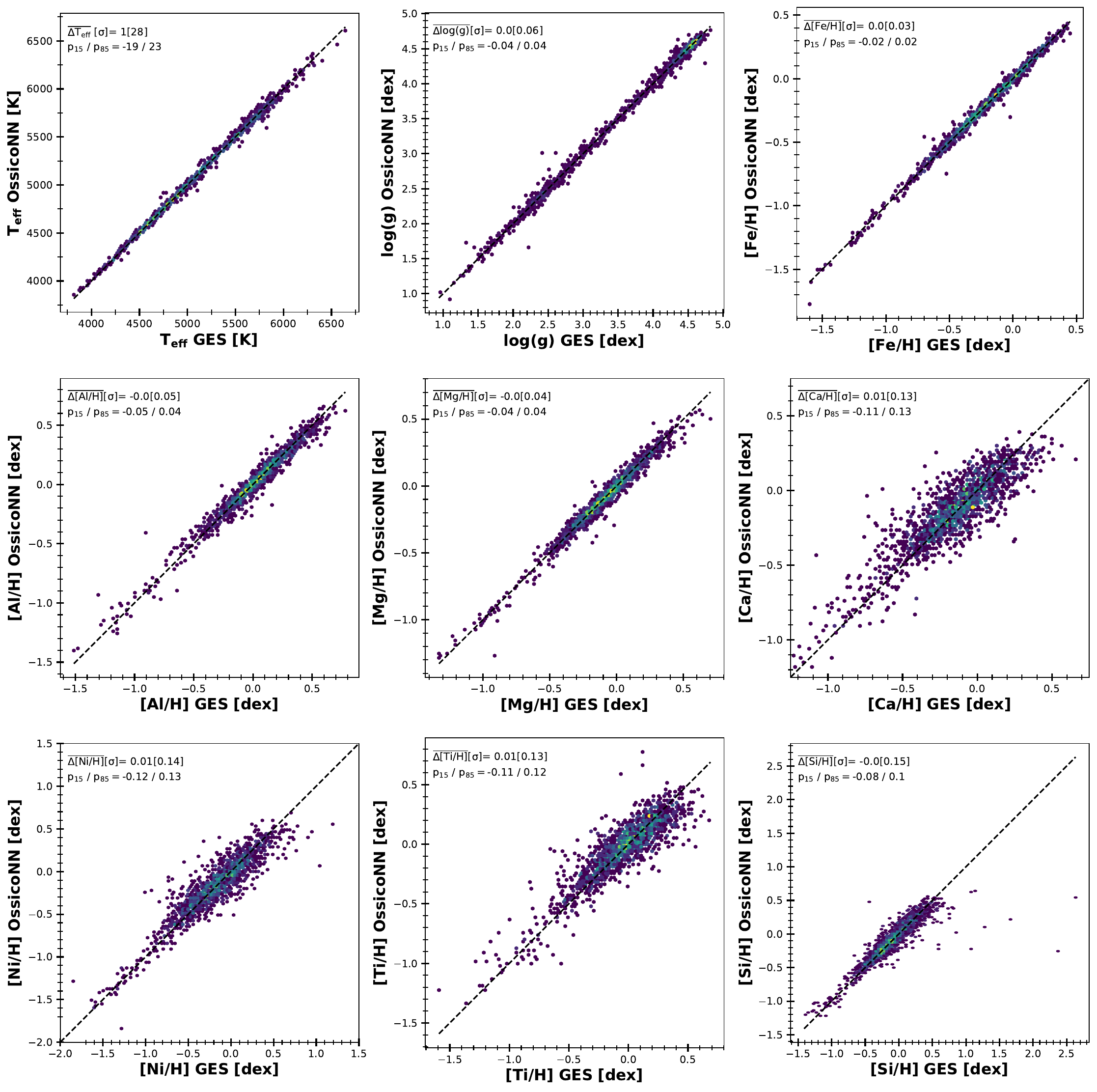}}
      \caption{Density distribution of parameters inferred by \texttt{OssicoNN} compared with the parameters given in GES for the test dataset. The average bias and the standard deviation, the 15th percentile, and  85th percentile of the residuals around the 1:1 relation are given in every panel. Dashed diagonal lines indicate the 1:1 relation.}
         \label{fig:param_ossiconn_ges_test_set}
   \end{figure*}

   \begin{figure*}
    \centering
   \resizebox{\linewidth}{!}
            {\includegraphics{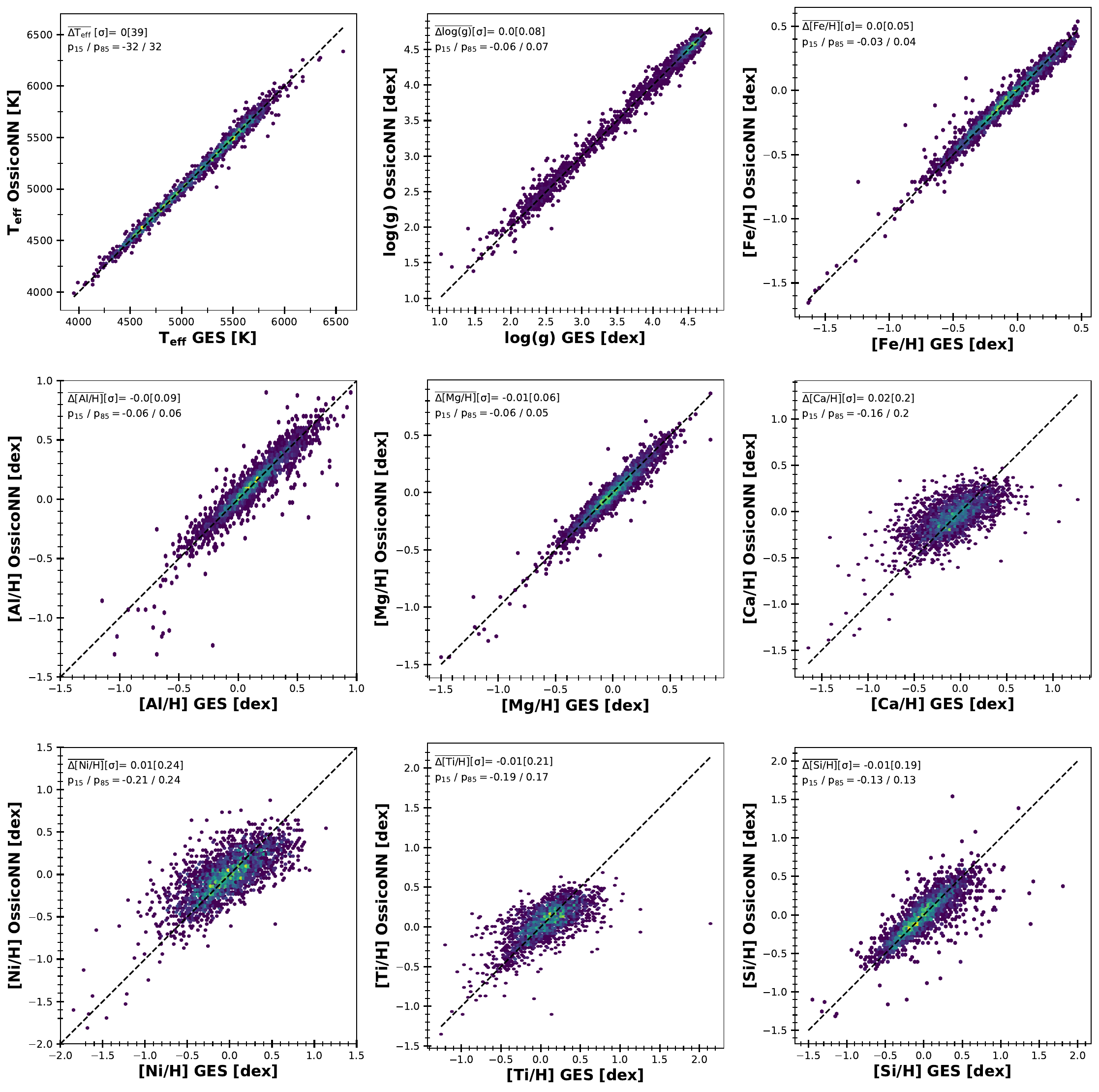}}
      \caption{Density distribution of parameters inferred by GES and \texttt{OssicoNN} for the noisy dataset. The average bias and the standard deviation, the 15th percentile, and  85th percentile of the residue results around the 1:1 relation are given in every panel. Dashed diagonal lines indicate the 1:1 relation. For the sake of clarity, we have excluded three stars from the [Si/H] quadrant that deviate significantly from the expected 1:1 relationship. These particular stars were initially predicted to have abundances of 3.1, 3.7, and 4.5 in the GES model. However, when using OssicoNN, their predicted abundances were notably different, measuring only -0.3, 0.2, and 0.2.}
         \label{fig:param_ossiconn_ges_noisy_set}
   \end{figure*}

\section{Results \label{sec4}}

\subsection{Accuracy of predictions}

\subsubsection{Test set predictions}

The neural network applied to the test set demonstrates no bias for either mode (1 K for temperature and >0.01 dex for both log gravity and abundances). The standard deviation of the residuals, which is indicative of the scatter in the direct comparison between the GES labels and OssicoNN predictions (Fig. \ref{fig:param_ossiconn_ges_test_set}), is relatively low. For the precision mode in particular, we obtain a standard deviation of 28K for temperature, 0.06 dex for gravity, 0.03 for metallicity, and ranges from 0.06 to 0.17 for the element abundances. The standard deviation consistently exceeds the absolute values of the difference between the 15th percentile p15 and 85th percentile p85 after median adjustment. This indicates a distribution that deviates from a Gaussian and is characterised by a narrow core with few outliers, exhibiting significantly divergent predictions. Moreover, the congruence of the mean and median across all parameters indicates that the predictions made by  OssicoNN are symmetrically distributed around the centre, where outliers are therefore few in number and are evenly distributed between overestimates and underestimates.

Table \ref{tab:param_ossiconn} and Fig. \ref{fig:param_ossiconn_ges_test_set} demonstrate that the alignment quality between predictions and labels for each parameter can be categorised into three levels: very high, high, and medium. Temperature, gravity, and metallicity predictions are notably accurate, exhibiting minimal scatter (translating into a low standard deviation for residues). The predictions for aluminum and magnesium abundances are also highly accurate, though they show a slightly larger scatter (about 0.05 dex). For other abundances, prediction becomes more challenging, with larger scatter (in the range of 0.1-0.2 dex). 

Figure \ref{fig:param_ossiconn_ges_test_set} highlights a small set of stars with very low metallicity (around $-$1.5 dex), which is predicted with remarkable accuracy despite the limited number of low-metallicity stars in the training set. We interpret this as a result of likelihood training, which compels the neural network to focus on less densely populated regions of the full parameter distribution. The distribution  of [Si/H] residuals deviates more strongly from a Gaussian, with a standard deviation that is 46\% greater than the 15th and 85th percentiles, while the difference between these statistics for [Ca/H], [Ni/H], and [Ti/H] is around 20\%. [Si/H] exhibits a narrow core and outliers for which the \texttt{OssicoNN} estimates are far from the GES labels. Overall, the agreement between the two estimates is good, although there are certain outliers where the \texttt{OssicoNN} predictions diverge from the GES labels.

\subsubsection{Noisy dataset predictions}

Figure \ref{fig:param_ossiconn_ges_noisy_set} shows the comparison between the input labels and the model predictions (using the Precision mode) on the noisy dataset. The median of the residuals remains strikingly consistent (0 K for temperature, 0 dex for gravity, 0 dex for metallicity, and between -0.02 dex and +0.01 dex for abundances). However, there is a noticeable uptick in the standard deviation (39 for temperature, 0.08 dex for gravity, 0.05 dex for metallicity, and between 0.06 dex and 0.24 dex for abundances). This increase is still minimal, considering that the median S/N drops from 55 to 24. Similar to the findings in the test set, the estimates for temperature, gravity, and metallicity align more closely with those from GES compared to the estimates for abundances (especially for [Ca/H], [Ni/H], [Ti/H], and [Si/H]).

Compared to Fast mode, Precision mode decreases the $u_{68}$ interval of the residuals between prediction and input labels (ranging between 0\% and 60\%) for various parameters. This improvement is particularly pronounced in noisy datasets. 

Using \texttt{OssicoNN}, we generated new estimates for the nine output parameters for the Catalogue dataset, totalling 49,858 stars. The table, available at the CDS, includes the object name in the GES DR5 catalogue (denoted as \textit{OBJ}), the nine parameters, along with their uncertainties (total, internal, and external), and a \textit{DATASET} entry indicating the dataset to which the star belonged (specifically whether or not it was part of the training set).

\subsection{Uncertainties}
\label{sec_uncertainty}

To investigate uncertainties, we applied the following restrictions to the Catalogue dataset: inferred temperatures ranging from 4000 to 6915 K and metallicities between -1.5 and 0.5 dex. Additionally, we filtered out stars with temperatures of below 4500 K and surface gravities exceeding $log (g)$ = 4. This step is designed to exclude stars falling outside the training range, resulting in a dataset comprising 42,738 stars from various datasets and a wide range of S/Ns. We refer to this dataset as the `Reduced Catalogue' dataset. 

This dataset contains the stars from the training set. This decision arises from the realisation that excluding stars would result in the omission of a large number of high-S/N spectra. However, we do not perceive this as a concern, as our primary interest lies in assessing uncertainties and not in the inferred values. Using only the test set does not result in significantly different quantitative trends. In the following, it is worth remembering that \texttt{OssicoNN} operates without knowledge of the estimated uncertainties associated with GES parameter labels. 

\begin{figure*}
    \centering
   \resizebox{\linewidth}{!}
            {\includegraphics{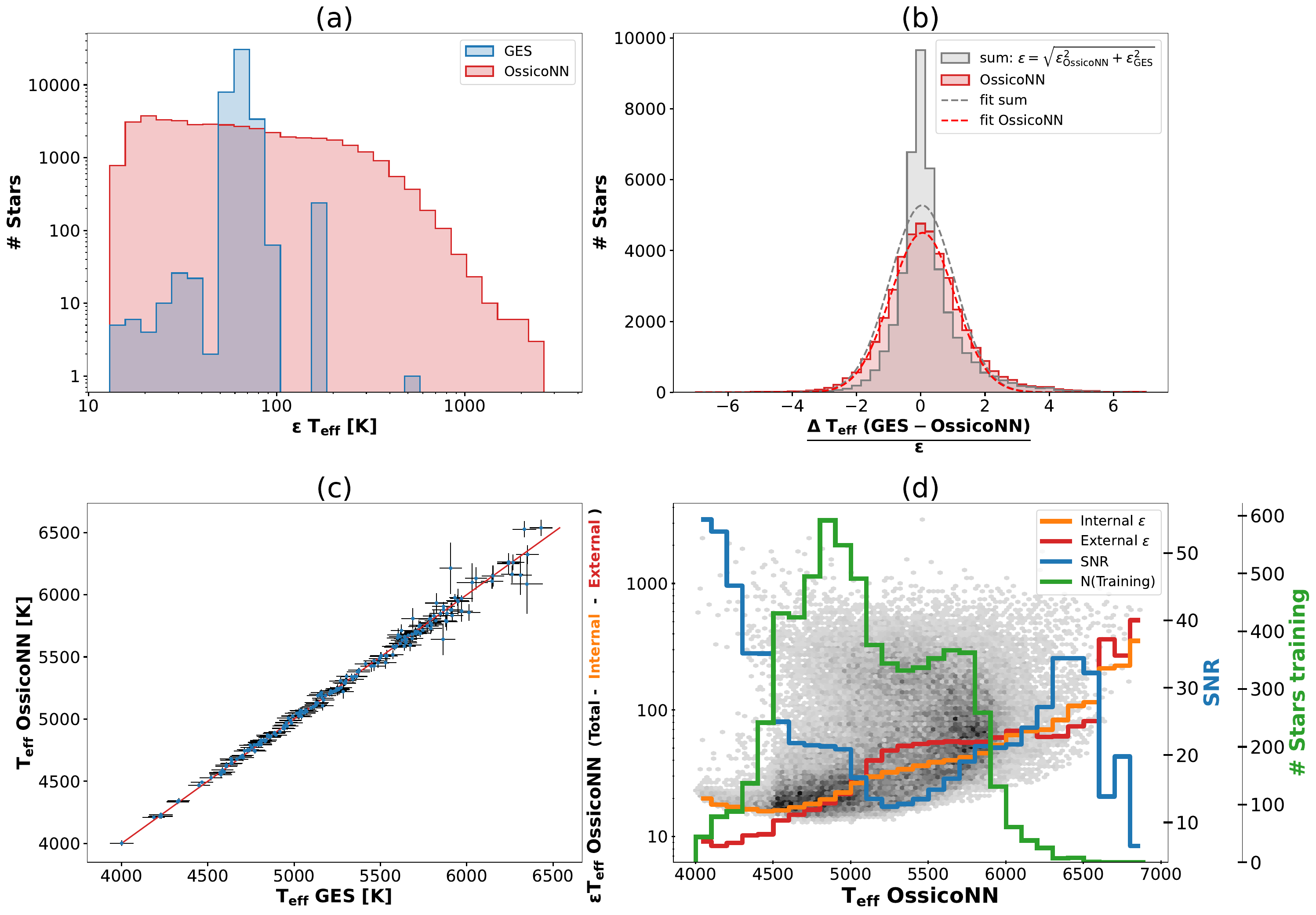}}
      \caption{Different plot for effective temperature uncertainties in the `Reduced Catalogue dataset' (totalling 42,738 stars). (a) Distribution of uncertainties in GES and \texttt{OssicoNN}. (b) Distribution of the residuals between \texttt{OssicoNN} and GES normalised with respect to the uncertainties of OssicoNN (red shadowed area) and  the sum in quadrature of GES and OssicoNN errors (grey shadowed area). Dashed lines represent the fit of the distribution with a Gaussian function with standard deviation $\sigma = 1.$   (c) Temperature estimates for 200 stars taken at random from the catalogue by GES and \texttt{OssicoNN} with their respective uncertainties. (d) Density distribution of total \texttt{OssicoNN} uncertainties with respect to inferred temperature. The red and orange lines represent the median external and internal uncertainties for each temperature bin. The blue curve and its scale on the right represent the median S/N per temperature bin.}
         \label{fig:uncertainties}
   \end{figure*}

Statistics detailing uncertainty distributions are provided in Table \ref{tab:uncertainties}. The overall temperature uncertainty distribution is shown in Fig. \ref{fig:uncertainties}a, and distributions for all other parameters can be found in Fig. \ref{fig:distribution_error}. Across all distributions, we  find that the uncertainty associated with each parameter is of the same order of magnitude as that estimated in GES. For example, in the case of temperature, the median of the uncertainty distribution is around 50 for \texttt{OssicoNN}, compared to approximately 64 for GES. The scale of the uncertainties in \texttt{OssicoNN} closely mirrors the errors observed between \texttt{OssicoNN} and GES, particularly for the Noisy dataset when \texttt{OssicoNN} is configured in Precision mode.

For temperature, surface gravity, and metallicity, we observe that the \texttt{OssicoNN} distribution is much wider, with a large tail and high uncertainties. For temperature, for example, this translates into a higher mean (101 K) in the \texttt{OssicoNN} distribution than the median and GES mean (64 K). An examination of the 670 stars with temperature uncertainties over 500 K shows an average S/N of five, highlighting the strong expected link between low S/N and high uncertainty estimations by the neural network.
For other abundances, the GES and OssicoNN distributions are similar, with the peak of the \texttt{OssicoNN} distribution occurring earlier than the GES peak, resulting in lower median and mean uncertainties for \texttt{OssicoNN}. Silicon is the exception, showing a larger tail and higher uncertainties ($p_{85}= 0.56$ dex). Silicon is also the least available parameter in the GES catalogue. It is assumed that when uncertainties were too high, the abundance was not recorded in the GES catalogue. In \texttt{OssicoNN}, this leads to very high uncertainties for silicon, making the measurement unreliable. 

The GES uncertainty distribution presents several distinct patterns (not visible in figures due to log scale). Firstly, the uncertainties in effective temperature, surface gravity, and metallicity exhibit discrete errors.
For other parameters with continuous distributions, we observe non-physical patterns, such as over-density and under-density strips, caused by the method used to compute the classical label. Conversely, \texttt{OssicoNN} does not reproduce these patterns.

Figure \ref{fig:uncertainties}b illustrates the distribution of normalised residuals for effective temperature, shown in red when considering only OssicoNN uncertainties and in grey when considering the quadratic sum of GES and \texttt{OssicoNN} uncertainties. Each distribution is fitted with a Gaussian with a fixed standard deviation of 1. According to the central limit theorem, this error-to-uncertainty ratio should follow a centred Gaussian distribution with a standard deviation of one if the relationship between GES and \texttt{OssicoNN} estimates is linearly homoscedastic and if the uncertainties are independently drawn from a normal distribution. Normalised residuals for all parameters are provided in Fig. \ref{fig:distribution_error_gauss}. Additionally, a similar figure using a reduced test set is also available. The Gaussian fits show that the errors are balanced, as indicated by the systematically zero means, regardless of the error and dataset considered. However, it is notable that some abundance distributions for the dataset of stars exhibit a tail towards negative values, with a very pronounced tail for metallicity and smaller tails for calcium, nickel, and titanium.

Based on the data, we propose that this tail results from two phenomena. Firstly, there is an overestimation of abundances because the training set contains few low-metal stars, leading to overestimation of these quantities (see Section \ref{sec4}). Secondly, as discussed in Section \ref{milky_way}, the metallicities of stars with very low S/N are underestimated in the GES dataset. These tails are absent in the test set (Fig. \ref{fig:distribution_error_gauss_test}) when the spectra are of good quality and fall within the training parameter space. When we normalise with respect to the sum in quadrature of OssicoNN and GES uncertainties, the distribution deviates from Gaussian. This deviation is expected, as the neural network is unaware of the uncertainties in the GES labels and assumes the labels provided to \texttt{OssicoNN} are absolutely true. 

Nevertheless, this does not mean there is no relationship between the estimates, labels, and their respective uncertainties. Figure \ref{fig:uncertainties}c shows the temperature estimates of GES and \texttt{OssicoNN} along with their uncertainties for a randomly selected subset of 200 stars (limited for clarity). Estimates are mutually compatible when accounting for measurement uncertainties as uncertainties almost universally overlap the 1:1 line. This concordance between estimates and labels is also illustrated by the large uncertainties attributed (by at least one of the estimators) when the pair (estimate and labels) deviates from the 1:1 relationship.
 
\begin{table*}
  \centering
    \caption{Statistics of  the distribution of the uncertainties in OssicoNN and GES for the reduced catalogue dataset.}
  \begin{tabular}{c c c c c c c c c c c c c c c c}
    \hline
    \hline
       & \multicolumn{5}{c}{$\mathbf{\epsilon  \ T_{eff}}$} & \multicolumn{5}{c}{$\mathbf{\epsilon \ log(g)}$} & \multicolumn{5}{c}{$\mathbf{\epsilon \ [Fe/H]}$}\\
    \hline
       & med & mean & std & p15 & p85 & med & mean & std & p15 & p85 & med & mean & std & p15 & p85\\
    \hline
    OssicoNN & 55 & 101 & 124 & 22 & 194 &
    0.08 & 0.13 & 0.15 & 0.04 & 0.25 & 
    0.05 & 0.07& 0.08 &  0.02 & 0.11\\
    GES & 64 & 64 & 11 & 53 & 66 & 
    0.15 & 0.15 & 0.02 & 0.14 & 0.16 & 
    0.07 & 0.08& 0.02 &  0.07 & 0.10\\
    \hline
    \hline
       & \multicolumn{5}{c}{$\mathbf{\epsilon \ [Al/H]}$} & \multicolumn{5}{c}{$\mathbf{\epsilon \ [Mg/H]}$} & \multicolumn{5}{c}{$\mathbf{\epsilon \ [Ca/H]}$}\\
    \hline
       & med & mean & std & p15 & p85 & med & mean & std & p15 & p85 & med & mean & std & p15 & p85\\
    \hline
    OssicoNN & 0.07 & 0.10 & 0.10 & 0.04 & 0.17 &
    0.06 & 0.12 & 0.25 & 0.04 & 0.20 
    & 0.16 & 0.21 & 0.14 & 0.10 & 0.34\\
    GES & 0.09 & 0.12 & 0.13 & 0.03 & 0.19 &
    0.10 & 0.14 & 0.16 & 0.05 & 0.22 & 
    0.20 & 0.24& 0.18 &  0.10 & 0.34\\
    \hline
    \hline
       & \multicolumn{5}{c}{$\mathbf{\epsilon \ [Ni/H]}$} & \multicolumn{5}{c}{$\mathbf{\epsilon \ [Ti/H]}$} & \multicolumn{5}{c}{$\mathbf{\epsilon \ [Si/H]}$}\\
    \hline
       & med & mean & std & p15 & p85 & med & mean & std & p15 & p85 & med & mean & std & p15 & p85\\
    \hline
    OssicoNN & 0.16 & 0.19 & 0.17 & 0.10 & 0.26 &
    0.17 & 0.19 & 0.18 & 0.09 & 0.26&
    0.15 & 0.36 & 0.72 & 0.09 & 0.56\\
    GES & 0.21 & 0.27 & 0.21 & 0.12 & 0.44 &
    0.20 & 0.25 & 1.1 & 0.11 & 0.34 & 
    0.11 & 0.16& 0.72 &  0.09 & 0.18\\
    \hline
  \end{tabular}
  \label{tab:uncertainties}
\end{table*}

Figure \ref{fig:uncertainties}d shows the distribution of inferred total uncertainties in temperature concerning the `Reduced Catalogue dataset'. Additionally, the figure illustrates the distribution of external (red) and internal (orange) errors in each temperature bin. For the corresponding bins, we also present the median S/N depicted in blue, alongside the distribution of the training set indicated in green. Comparable visualisations for additional parameters can be found in Fig. \ref{fig:uncertainties_alea_epistemic}.

First, for high uncertainties characterised by $\epsilon T_{eff}>500 K$, the errors exhibit a uniform distribution. This uniform distribution of high uncertainties is consistently observed across other parameters as well. Subsequently, we can decompose the total error into external and internal components. The external error demonstrates a strong anticorrelation with the S/N: it is inversely related, such that high S/N values correspond to low external errors, and conversely, low median S/N values result in high errors assigned by the neural network. Specifically for temperature, the external error reaches a minimum of 8 K around 4200K when the median S/N exceeds 50, and it escalates to a maximum of 513K at 6900K when the median S/N falls below 10. This trend is consistently observed across all parameters.

The internal error manifests as a convex function, initially valued at 20 K, decreasing to 16 K at 4500 K, then rising to 83 K at 6350 K, and ultimately surging to 352 K at 6850 K. This error distribution is influenced, among other factors, by the distribution of the training set: a higher density of the training set at a specific temperature correlates with a lower external error. This relationship is particularly evident in the abundance parameters, where the minimum internal uncertainty coincides with the maximum training set distribution. Additionally, this function likely depends on the neural network's capability to optimize hyperparameters and accurately map the parameter space to the latent space, effectively performing a precise fit. It is anticipated that the contribution of this factor is minimal at the centre of the parameter distribution and maximal at the boundaries, though further investigation and quantification are necessary.

Nonetheless, this internal uncertainty is modulated by the S/N, as a high S/N suggests a higher proportion of training stars within a bin relative to the total star population in that bin, as well as an enhanced capability to extract useful information due to reduced noise. However, this analysis does not hold for temperature, where we would anticipate a minimum internal error at 4750K.
This unexpected behaviour is attributed to the degeneracy (refer to Figures \ref{fig:distri_pearson_resume} and \ref{fig:kiel_diag}) between the main sequence and the red giant branch, which is obscured by marginalisation along the temperature axis. \texttt{OssicoNN} reflects this degeneracy as a high internal error. To verify this, we examined Fig. \ref{fig:uncertainties}d to ensure that the stars of the giant branch and main sequence were not divided into distinct zones (one with low uncertainties and the other with high uncertainties) and observed an overlap between the populations. The other parameters do not exhibit such degeneracy and thus more accurately reflect the training set distribution.

In summary, both internal and external errors exhibit behaviours consistent with the uncertainties that have been included and listed in Section \ref{sec_uncertainty_theory}, namely dependencies on S/N, on the training distribution, potential degeneracies, and so on. Similar trends can be seen for all other parameters in Fig. \ref{fig:uncertainties_alea_epistemic}. The external error demonstrates a robust correlation with the S/N of the spectrum, with higher S/N associated with lower external error, and vice versa. Internal uncertainties are represented by a function that reflects the distribution observed in the training set, the geometry in parameter space, and the challenges due to the non-deterministic relations between spectra and parameters. In Gaussian training distribution with non-degeneracy, this distribution tends to be convex, with its minimum coinciding with the peak density of the training set.

\subsection{Interpreting the network with saliency maps}
\label{sec_saliency}
Given the nature of stellar physics, not every combination of stellar parameters is feasible, resulting in correlations between certain parameters. For instance, metal abundances exhibit correlations, and so, for example, a high iron content tends to coincide with a high aluminium abundance. These correlations are inherently present in our training dataset, along with any correlations arising from observational constraints such as instrumental ones, as well as choices made regarding the study of particular populations. During the neural network training process, access to these parameters enables the network to derive hyperparameters based on these correlations rather than solely from spectral data. While this approach may yield satisfactory results for many stars, it may falter for more unconventional cases. 

\begin{figure*}
    \centering
   \includegraphics[width=0.9\linewidth]{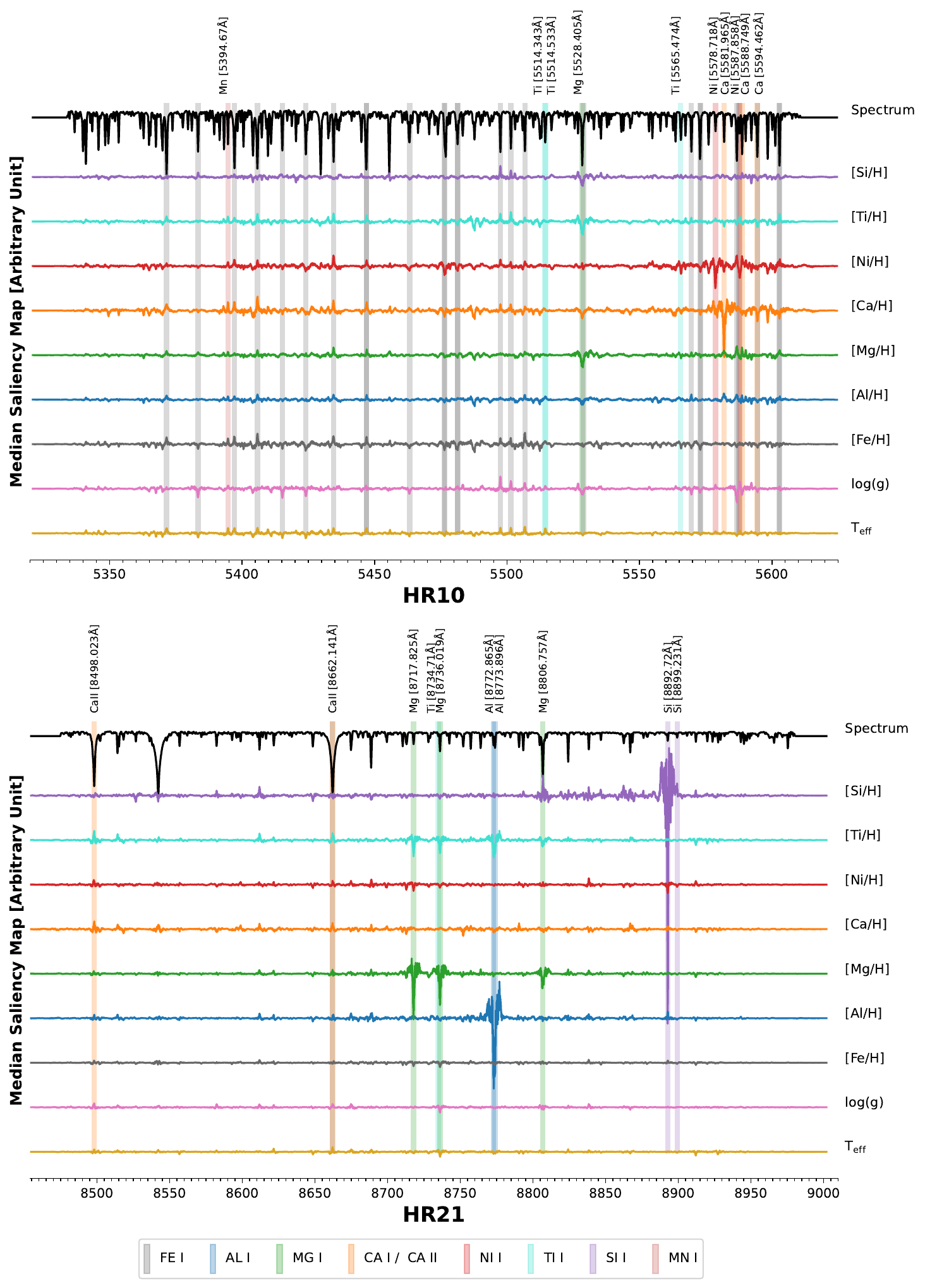}
    \caption{Median saliency map across all test and noisy datasets. This is obtained by averaging the saliency maps for each parameter. The represented spectrum corresponds to the median of the test and noisy datasets spectra. The primary theoretical lines corresponding to a peak in the absolute value of the saliency map are indicated by vertical lines. The displayed iron lines are among the 15 most significant peaks for the iron saliency map.}
    \label{fig:saliency_map}
    \end{figure*}

To verify what features of the training set the network learnt from, we can examine the saliency map. A saliency map identifies the most influential regions or features of an input image that contribute to the network's output. Briefly, this method computes the gradient of the output score concerning the input image and utilises this gradient information (saliency map = $\frac{\partial \mathrm{param}}{\partial \mathrm{spectra}}$) to pinpoint the regions of the input image exerting the greatest influence on the output.
The saliency map is computed for all features and spectra within the test dataset and is then averaged across all spectra. Figure \ref{fig:saliency_map} illustrates the saliency map for all parameters. We then compute the absolute values of the saliency maps and determine the main peaks and their associated wavelengths. Using \textsc{scikit-learn} \cite{scikit-learn} and the \texttt{NearestNeighbors} function, we match the wavelengths of the peaks  with the theoretical lines in a catalogue of atomic lines pertinent to this domain \citep{Heiter:2021}.

As evident from Fig. \ref{fig:saliency_map}, temperature ($T_{\text{eff}}$) and surface gravity ($\log g$) are typically not reliant on individual lines but rather on the entire spectrum. More specifically, the temperature estimate comprises a combination of several lines with approximately equal weights, including those associated with magnesium I, iron I, manganese I, titanium I, and calcium II.

Concerning surface gravity, \texttt{OssicoNN} primarily derives pertinent information from HR10, focusing particularly on iron lines (with the ones at $5497.516\AA$ and  $5586.80\AA$ being the most crucial), as well as lines related to magnesium and calcium.

When examining the abundances, our expectation is for them to be deduced mostly from the lines associated with the respective elements. The saliency map for [Fe/H] captures information from numerous spectral lines. Specifically, the map identifies the  15 main peaks associated with 18 theoretical lines, with some peaks corresponding to multiple spectral lines due to the pair-matching algorithm erroneously assigning two theoretical lines to the same saliency map peak. Each of these peaks can be associated with one or more theoretical iron lines.

Similarly, when averaging over all spectra, the saliency map for aluminium, magnesium, and silicon is predominantly influenced by a select few lines corresponding to each element. Specifically, aluminium is primarily determined by the doublet at $8772.865 \AA$ and $8773.896 \AA$, magnesium by the lines at $8717.825 \AA$, $8736.019 \AA$, $8806.757 \AA,$ and $5528.405 \AA$, and silicon by the line at $8892.720 \AA$, although a handful of additional lines from other elements may contribute marginally.

A second group of elements, including nickel and calcium, primarily relies on their respective spectral lines ($5581.965 \AA$ for calcium and $5578.718 \AA$ for nickel), as well as the silicon line at $8892.720 \AA$. The saliency map also gives some importance to secondary contributions from other lines of the element under consideration ($5587.858 \AA$ for nickel and $5594.462 \AA$ for calcium), as well as aforementioned lines from other elements (aluminium, magnesium and silicon).

Titanium poses a significant challenge as it is primarily determined by several lines originating from other elements, notably the ones mentioned above, and only to a limited extent by titanium-specific lines such as $5490.148\AA$, $5514.343\AA$, and $5514.533\AA$.

Based on these findings, it appears that aluminium, magnesium, and silicon are determined independently of the other elements, while calcium and nickel show slight correlations with silicon. Titanium, on the other hand, seems to be influenced by a combination of other elements and may exhibit correlations with the abundances of other elements.

The saliency maps show that the information content of each spectral region is never zero. This effect is due to the use of convolutional layers with a large kernel, which aggregates information from a substantial number of pixels simultaneously, leading to a smoothing effect in the saliency map calculation.

Additionally, we examined the impact of sampling parameters based on their values. Spectra were sampled according to their aluminium abundance ($-1<[Al/H]; -1\le[Al/H]<-0.5; -0.5\le[Al/H]<0; 0\le[Al/H]<0.5 ; [Al/H]>0.5$). At higher abundances, reliance on the doublet $8772.865\AA$ and $8773.896\AA$ is evident. 

\section{Astrophysical validation \label{sect:astrophysical_validation}}

To further evaluate the precision and the accuracy of \texttt{OssicoNN}, we compared the parameters obtained by the pipeline with those derived from the GES workflow for three selected samples (field stars, star clusters, and benchmark stars) using some of the astrophysical tests used in \cite{Hourihane:2023} to check the quality of the data.

\subsection{Field stars}
\label{milky_way}

From the Catalogue dataset, we selected stars that belong to the Milky Way field using the keyword GES\_TYPE = `GE\_MW', where temperature, surface gravity, and metallicity were estimated by the GES pipeline. This selection includes stars from both the thin and thick Galactic disc, along with a small fraction of halo stars, totalling 44,756 stars.

\begin{figure*}
    \centering
    \resizebox{0.85\linewidth}{!}{\includegraphics[angle=0]{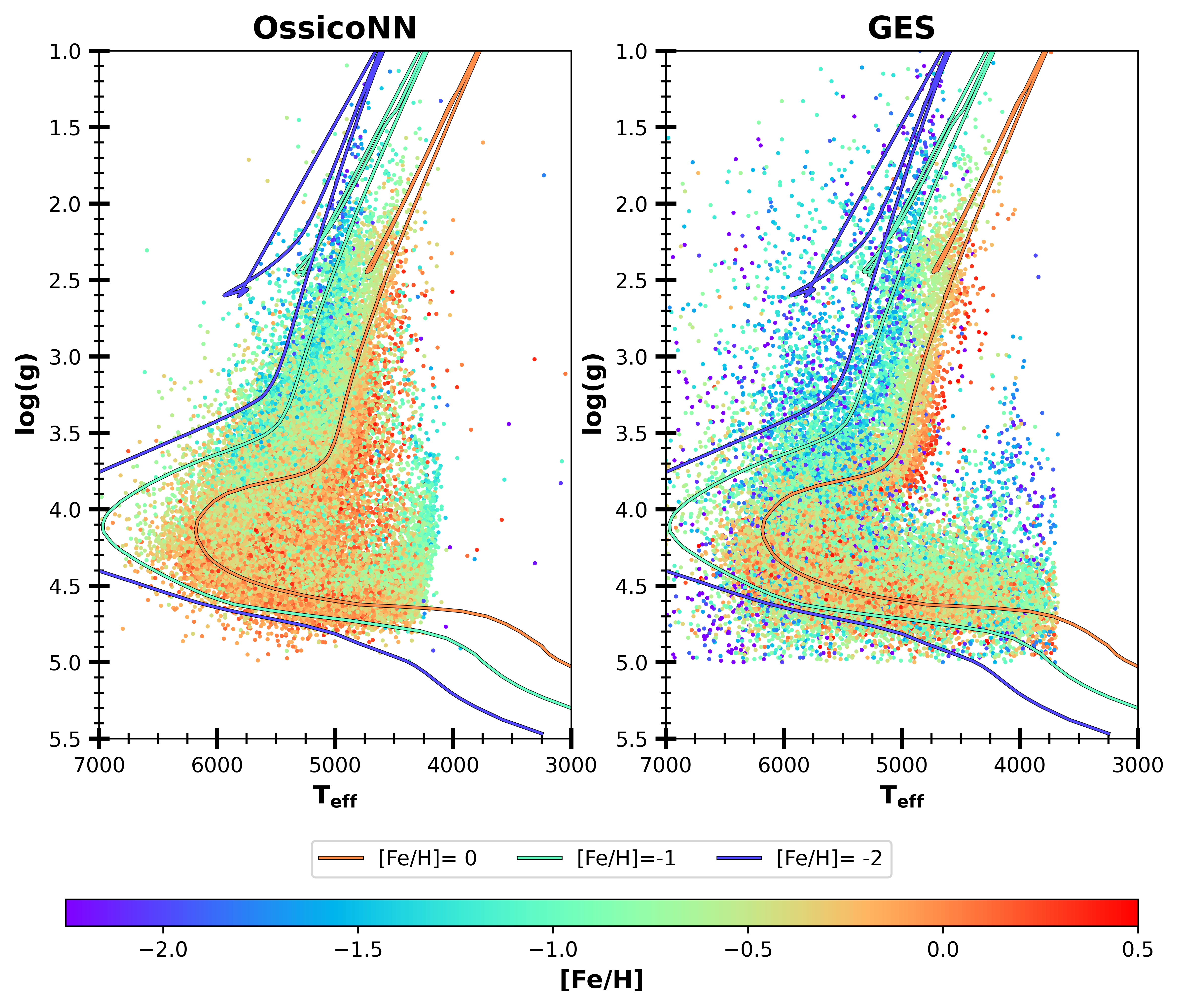}}
    \caption{Kiel diagram: log~g versus effective temperature  colour coded according to metallicity for stars in the Milky Way field as derived using \texttt{OssicoNN} astrophysical parameters (left panel) and the GES recommended values (right panel). Superimposed are three isochrones for an age of 5 Gyr and three metallicities ([Fe/H]=0: orange, [Fe/H=1]: green, [Fe/H=1]: blue) with colours that match the colormap). generated using PARSEC version 1.2S (\cite{parsec_bressan}, \cite{parsec_chen}). Stars were selected from the Catalogue dataset where `GES\_TYPE = GES\_MW', with temperature, surface gravity, and metallicity derived by the classical pipelines, totaling 44,756 stars.}
    \label{fig:kiel_diag}
\end{figure*}

Figure \ref{fig:kiel_diag} shows the Kiel diagrams of this sample created with the \texttt{OssicoNN} (left panel) and GES (right panel) parameters. This diagram depicts the relationship between surface gravity, effective temperature, and metallicity, providing insight into stellar evolution. Comparing the diagrams reveals that \texttt{OssicoNN} accurately reproduces the structure observed in GES data. Notably, we can observe the presence of a well-populated main sequence (MS) region characterised by surface gravity of between 4.8 and 4.3, temperatures spanning from 4200 K to about 7000 K, and solar metallicity. 
Low-temperature stars (around 4000K) and surface gravity higher than 3.5, thus located about the MS,  are expected to pre-main sequence stars.
The red giant branch (RGB, with T$_{\rm eff}$ from about 4500 to 5500 K and log~g$<$3.5) is also well populated by stars of different metallicity. As expected, the more metal-rich stars at any given log~g are the cooler ones. The trend is well reproduced with the  \texttt{OssicoNN} parameters, in which the scatter is reduced ---in particular on the left side of the  RGB for the lower metallicity giant stars. 

However, \texttt{OssicoNN} struggles to accurately replicate some stars with very low metallicity below $[Fe/H] = -2$, particularly those with high temperature and gravity (around 6500 K and 4.7 dex), or low temperature and medium gravity (around 4000 K and 4 dex). These stars are mainly characterised by spectra with S/N ranging between 5 and 10. Furthermore, as these stars were not included in the training dataset, \texttt{OssicoNN} did not learn to effectively distinguish between them.  Moreover, we argue that the parameters measured for these stars in GES are not correct: for the upper MS stars, the low metallicities obtained are due to degeneracy between high temperature and low metallicity, while for the lower MS and PMS,  this is to the presence of molecular bands that do not allow the stellar continuum to be correctly estimated. From a chemical evolution point of view, we do not expect to have upper MS (thus massive and young) stars in the disc with such low metallicity ([Fe/H]$<$-2). Along the same line of argument, the cool PMS stars representing the late stages of Galactic chemical evolution are expected to have at least solar metallicity \citep[see e.g.][]{Magrini2023A&A...669A.119M}.

On the other hand, the uncertainties assigned by \texttt{OssicoNN} to these stars are relatively low, particularly the internal uncertainty, despite the higher external uncertainty caused by the low S/N. This merits a significant level of confidence in the predictions made by \texttt{OssicoNN}. Similarly, when examining the placement of these stars in the Kiel diagram using \texttt{OssicoNN}, we observe a good alignment with the overall trend seen among stars with similar parameters. Whether any inaccuracies in our assessment of \texttt{OssicoNN} are due to limitations in the training set or errors in our evaluation of the GES pipeline because of the low S/N remains uncertain. Nonetheless, our analysis leans towards the latter hypothesis, highlighting once again the remarkable effectiveness of the neural network in the low-S/N domain.

\begin{figure*}
    \centering
    \resizebox{0.85\linewidth}{!}{\includegraphics[angle=0]{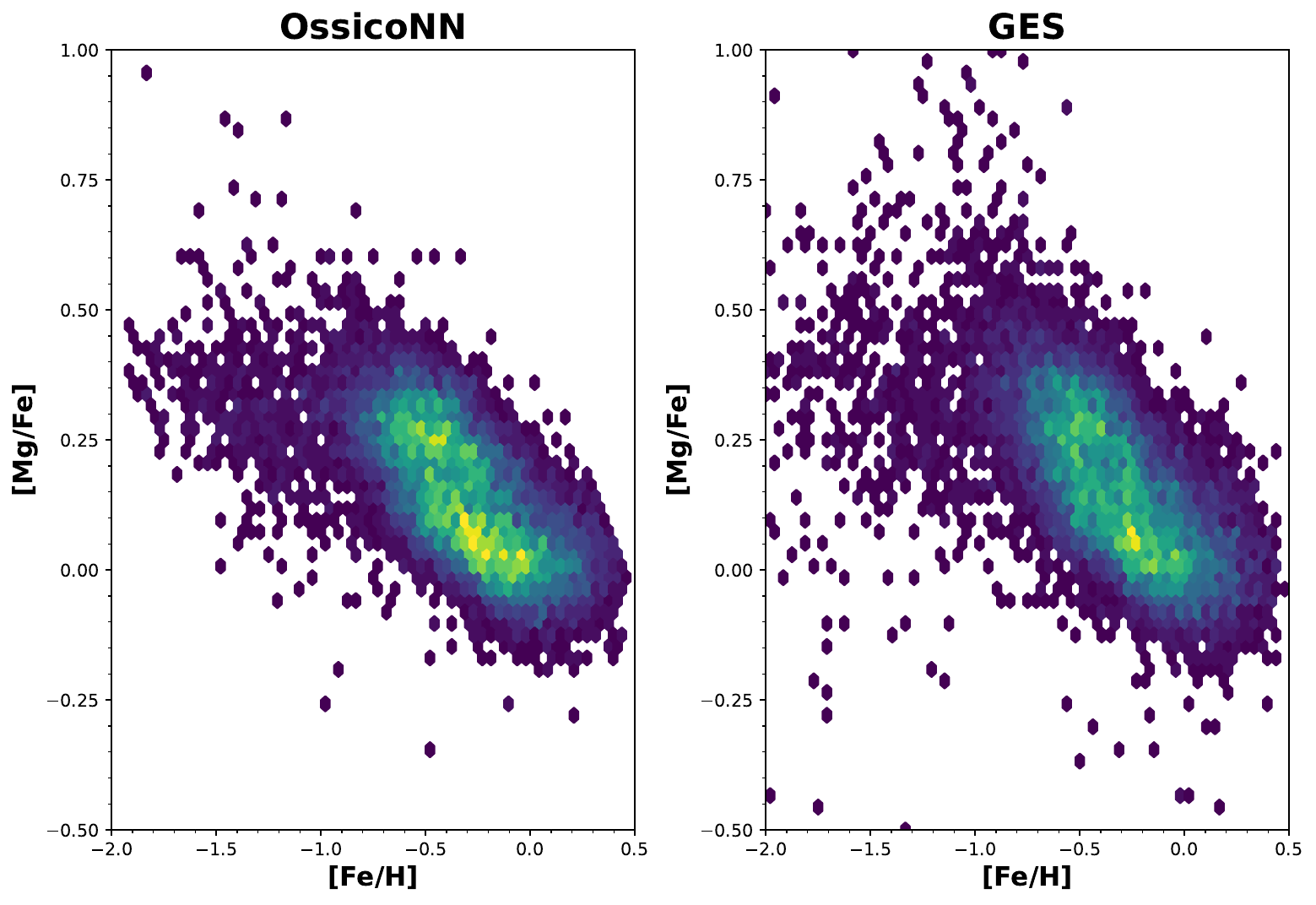}}
    \caption{Abundance ratio of magnesium to iron ([Mg/Fe]) versus metallicity ([Fe/H]) for  stars in the Milky Way field as derived using \texttt{OssicoNN} astrophysical parameters (left panel) and the GES recommended values (right). Stars were selected from the Catalogue dataset where `GES\_TYPE = GES\_MW', with [Fe/H] and [Mg/H] derived by the classical pipelines, totalling 11,909 stars.}
    \label{fig:thick_thin}
\end{figure*}

Figure \ref{fig:thick_thin} shows the distributions of [Mg/Fe] versus [Fe/H] for all the stars with S/N greater than 25 obtained using the parameters derived with \texttt{OssicoNN} (on the left) and the GES workflow (on the right). To perform this task, it is essential to compare the same stars, requiring that both [Fe/H] and [Mg/H] be determined by the GES pipeline. As a result, the figure includes a total of 11,909 stars. This plot is classically used to separate the two disc populations \citep[see e.g.][]{Fuhrmann1998A&A...338..161F, Bensby2003A&A...410..527B, Hayden2017A&A...608L...1H}: at a given metallicity, thick-disc stars are expected to have higher [Mg/Fe] due to their different star formation history. 
In Fig.\ref{fig:thick_thin}, the thin and thick disc are better separated when the \texttt{OssicoNN} abundances are used (left panel) than when the original GES ones are employed. This suggests that the abundances derived with our neural network are more precise than those obtained with classical analysis, improving the analysis of the low-S/N spectra and  consequently improving the separation and reducing the scatter. 

\subsection{Star clusters}

Star clusters are often used for the calibration and the validation of pipelines used for the analysis of stellar spectra. This is because they are composed of coeval stars with known age that span a large range of effective temperatures and surface gravities and, depending on the elements and on the type of cluster, are chemically homogeneous within 0.02-0.05 dex 
\citep{Bovy:2016, Poovelil:2020}.
We note that, instead,  globular clusters might show a large spread in some abundances, which is related to anti-correlations between the abundances of pairs of elements \citep[see e.g.][]{Pancino:2017b}.

Here, we compare the astrophysical parameters from \texttt{OssicoNN} and GES for three open clusters NGC 2243, NGC 2420, and Br 32, and three globular clusters NGC 1904, NGC 362, and NGC 2808. We chose these three open clusters because they include a significant number of stars observed with the HR10 and HR21 setups and the globular clusters because they allow us to test the pipeline within  a range of metallicities (from [Fe/H] -1.55 to  -0.82~dex) that is poorly covered in the training sample.

 To identify the stars of a specific cluster in the GES catalogue, we used the keywords GES\_FLD=CLUSTER\_NAME and MEM3D $>$ 0.9. The first keyword selects all the stars that were included in the GES target list as possible cluster members, while the latter identifies stars with a probability of greater than 90\% of belonging to the cluster on the basis of \textit{Gaia} proper motions and GES radial velocities. 

The distributions of the stellar parameters and the abundances for the three open clusters are shown in Figs. \ref{fig:open_cluster_br32}, \ref{fig:open_cluster_NGC2243}, and \ref{fig:open_cluster_NGC2420}. The figures reveal that the predicted surface temperature and gravity exhibit a distribution mirroring that of GES. On average, the standard deviation of the chemical-abundance estimates by OssicoNN in open clusters is reduced by 33\%, though this varies significantly from one abundance to another. For example, the dispersion for magnesium does not decrease, while the magnesium dispersion is reduced by 70\%. For all elements, OssicoNN estimates show fewer outliers and a denser core. However, the clear reduction in the spread for titanium might be linked to the weak influence of titanium lines in determining abundance (see Section \ref{sec_saliency}), leading to a less accurate estimate of titanium diversity compared to other elements. As such, extreme caution should be exercised when drawing conclusions regarding titanium, particularly when attempting to identify behaviour diverging from the norm.

We also display the Kiel diagram of each cluster with the isochrone corresponding to the age and metallicity derived by \citet{cantat-gaudin:2020} and reported in \citet{Randich:2022}, which were generated using PARSEC version 1.2S (\cite{parsec_bressan}, \cite{parsec_chen})\footnote{http://stev.oapd.inaf.it/cgi-bin/cmd}. For each cluster, we calculated the distance between the stars and the isochrone using a KDTree. The RMSE (root mean square error) was then calculated. The RMSE of \texttt{OssicoNN} is reduced by 10\% for NGC 2243, remains similar for NGC420, and increases by 25\% for Br32. Overall, the stars of the subgiant branch, the RGB, and the red giant clump are well positioned in relation to the isochrone by both GES and \texttt{OssicoNN}.

The differences between GES and \texttt{OssicoNN} are most pronounced in the main sequence and the main sequence turn-off regions. In NGC 2243, the turn-off is better defined by \texttt{OssicoNN}, while in Br32, the GES estimates are closer to the isochrone. In the more complex case of NGC2420, there is a sequence of stars that appear to belong to the main sequence, where GES predicts a decrease in gravity starting from $T_{eff} \approx 6600$K. Consequently, this star matches the isochrone section corresponding to the subgiant branch. If we exclude the subgiant branch from the isochrone and the other stars belonging to this branch, we obtain an RMSE of 35 for GES and 32 for \texttt{OssicoNN}, corresponding to a 10\% reduction. Another issue observed in NGC2243 is the relationship between temperature and metallicity. In NGC2420, some high-temperature stars (particularly in GES) exhibit non-physical metallicities (too low compared to the rest of the open cluster). The presence of this trend in both labels and predictions (without being in the training set) indicates a potential issue in the spectra.

Globular clusters exhibit distinctive anticorrelations in some of their abundance ratios, a phenomenon that is likely indicative of chemical enrichment resulting from the nucleosynthetic processes undergone by preceding generations of stars. The manifestation of this chemical signature is particularly pronounced in the relationship between magnesium (Mg) and aluminium (Al) within these clusters \citep{Pancino:2017b}.

\begin{figure}
   \centering
   \includegraphics[width=\hsize]{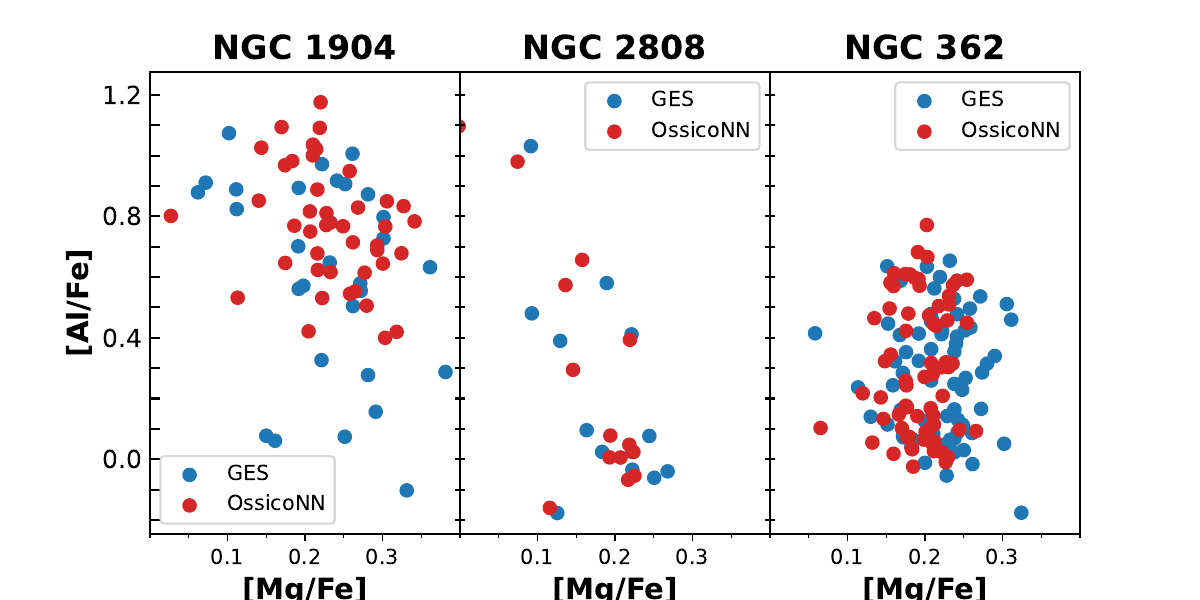}
      \caption{Anticorrelation of chemical abundances [Al/Fe] and [Mg/Fe] observed in globular clusters when parameters are inferred by \texttt{OssicoNN} (red) or GES (blue).}
         \label{fig:GC_al_Mg}
   \end{figure}

These clusters serve as a robust benchmark for evaluating the neural network's efficacy in replicating astrophysical phenomena, specifically through its interpretation of the Mg-Al relationship. Figure \ref{fig:GC_al_Mg} shows the relation between [Mg/Fe] and [Al/Fe] abundances for the three globular clusters included in this test. The figure demonstrates the neural network's adeptness in faithfully reproducing the anticipated anticorrelation between Mg and Al. We measured the Pearson coefficients between these two parameters across all clusters and found a good correspondence between the GES pipeline estimates and the \texttt{OssicoNN} estimates, with an average difference of 0.1, emphasising the network's potential for enhanced precision in astrophysical analyses.

\subsection{Benchmark stars}

\begin{figure}
    \resizebox{\linewidth}{!}{\includegraphics[angle=0]{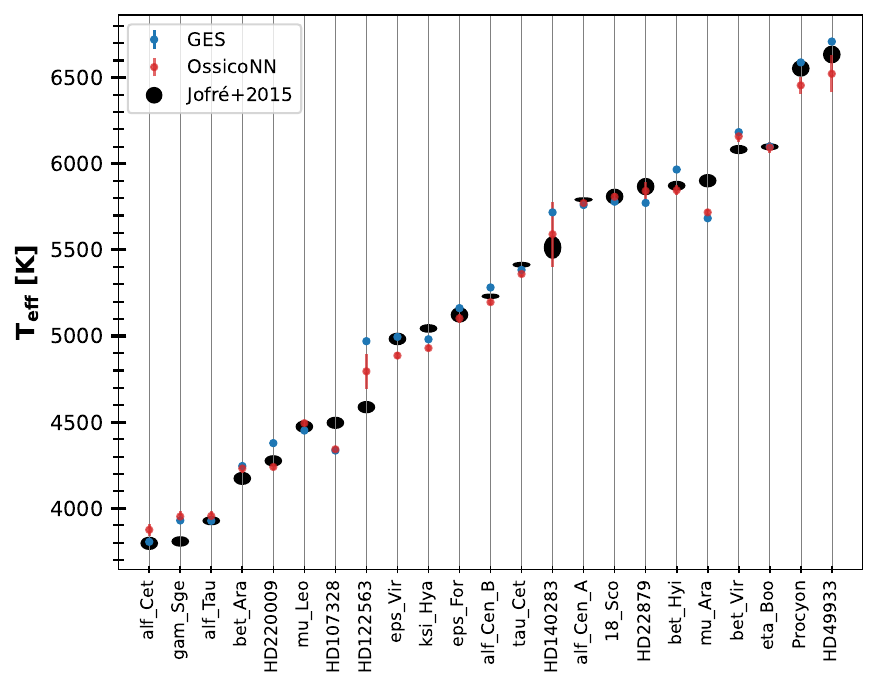}}
    \caption{Effective temperature estimations for selected Benchmark stars using the GES pipeline, the \texttt{OssicoNN} neural network, and \cite{jofre:2015}, who combine spectral and spectral-independent analyses. The size of the Jofré+2015 dots takes into account the measurement uncertainty.}
    \label{fig:teff_benchmark}
\end{figure}

\begin{figure*}[]
\centering
\includegraphics[width=\hsize]{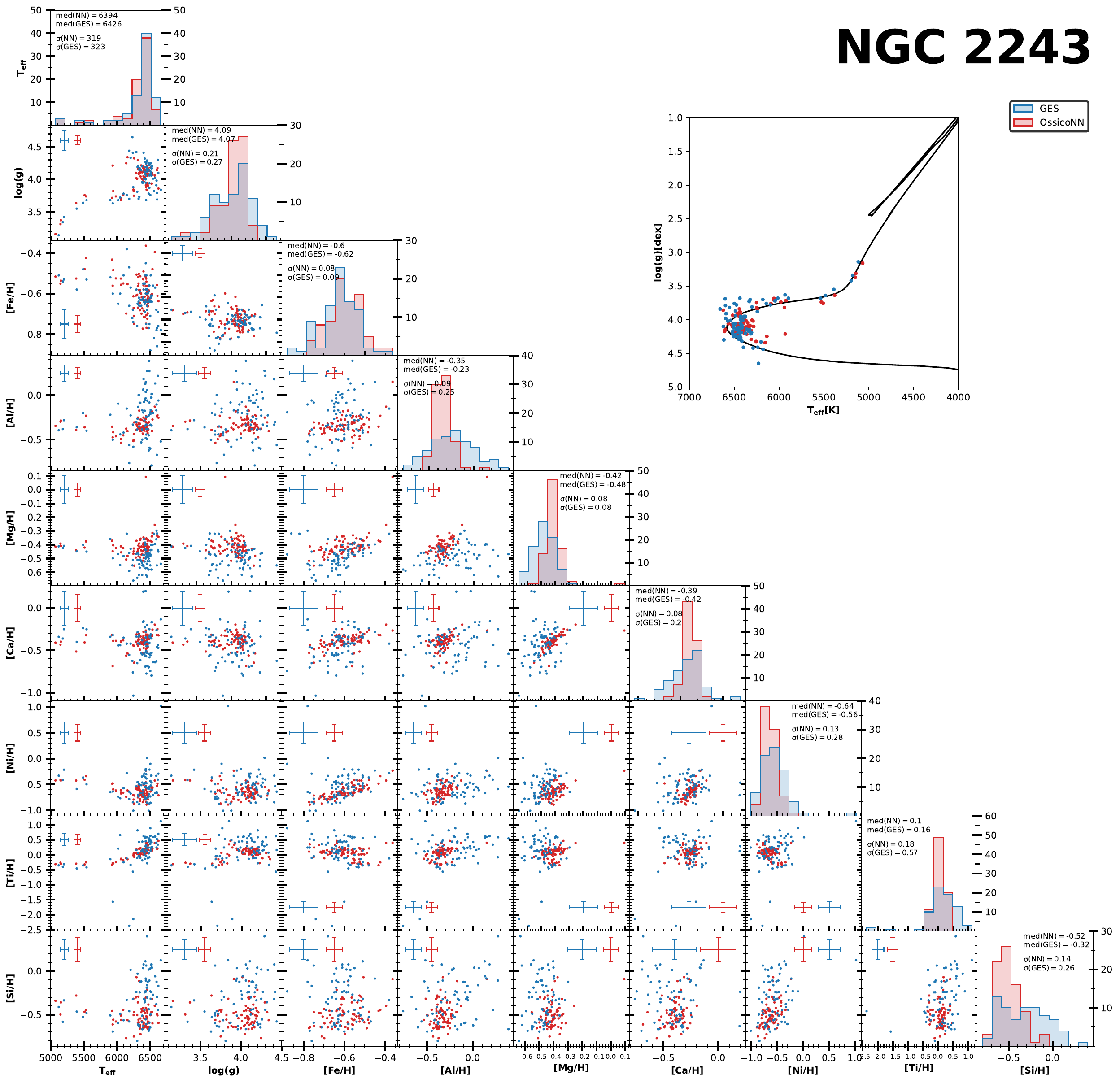}
  \caption{Parameters and abundances of stars belonging to the NGC2243 cluster using the GES pipeline and the OssicoNN neural network. The number of GES stars varies across quadrants since it depends on how many stars could be successfully processed through the classical pipelines for the given set of parameters. We also display the Kiel diagram for and the isochrone corresponding to the age and metallicity of the cluster measured by \cite{cantat-gaudin:2020}. The isochrone was generated using PARSEC version 1.2S (\cite{parsec_bressan}, \cite{parsec_chen}).}
     \label{fig:open_cluster_NGC2243}
\end{figure*}

In this section, we present a comparative analysis between the OssicoNN and GES predictions and measurements obtained from alternative studies, focusing on benchmark stars. Figure \ref{fig:teff_benchmark} provides a visual representation of the disparity between the GES/\texttt{OssicoNN} predictions and external estimates by \cite{jofre:2015} for temperature. Comparisons for the other parameters are shown in Fig.\ref{fig:benchmark_all}. The depicted figures unequivocally illustrate the accuracy of the predictions.

Three groups emerge. The first consists of Alf Cet, Gam Sge, Alf Tau, Bet Ara, and HD220009. These five stars exhibit erroneous predictions for several parameters, albeit with large uncertainties, highlighting the importance of cINN in assessing the quality of inference. This discrepancy is likely due to these low-temperature, low-gravity stars not being present in the training set, leading to incomplete inference on most parameters, except for temperature and metallicity. 
Similarly, the second group, comprising HD122563 and HD140283, which are characterised by low metallicity ($[Fe/H]<-2$, outside the training range), shows completely inaccurate inferences by \texttt{OssicoNN}, albeit with large uncertainties.
The third group comprises the remaining stars, which are predicted quite well with a precision comparable to that of GES. For instance, the average difference between the \texttt{OssicoNN} predictions and the estimates of the reference for temperature is -52 K on average, compared to -15K between GES and the reference, with a maximum difference of 186 K for \texttt{OssicoNN} versus 183 K for GES. Concerning gravity, the average difference between the reference and \texttt{OssicoNN}  is 0.04, whereas it is -0.04 for GES. Abundance predictions are similarly accurate, with an average deviation between the neural network and the reference ranging from 0.02 dex for silicon to 0.06 dex for titanium. This deviation is slightly higher for \texttt{OssicoNN} than for GES, except for calcium and silicon, where the \texttt{OssicoNN} prediction shows closer agreement with the reference. 

These findings underscore the data-driven nature of the neural network, highlighting that it lacks sufficient encapsulation of physical principles to differentiate itself from GES in favour of a more fundamentally grounded estimate.
The analysis reveals that estimates of both surface gravity and abundance are less precise for low-temperature stars particularly those falling within the M star category. This observation underscores a potential limitation in the predictive capacity of the model, particularly in scenarios involving low-temperature stars.

    \section{Discussion}
    \label{sec_discussion}

Many of the limitations stem from the training dataset's insufficient coverage of the full parameter space.
The challenge arises from the necessity to retain only stars with the complete set of nine parameters for training, which significantly diminishes the available pool of stars. To address this issue, we experimented with estimating values for spectra missing up to two parameters (excluding temperature, gravity, and metallicity). This estimation involved interpolating the absent abundances based on metallicity (see Appendix \ref{Sec:Agu} for more details). The rationale behind this approach is to save stars that lack only a few parameters, while the majority are well inferred by GES. By implementing this technique, we effectively doubled the size of the training dataset and substantially augmented its diversity. Consequently, this enhancement led to improvements in overall prediction accuracy and rectified issues observed in the Kiel diagram or with benchmark stars of M-type and low metallicity. However, by calculating the Pearson product--moment correlation coefficients for the GES and OssicoNN estimates with and without augmentation, we observed a discrepancy. Specifically, the coefficients for silicon and titanium showed a divergence when comparing GES and classic \texttt{OssicoNN} to augmented \texttt{OssicoNN}. This divergence is particularly severe for silicon, which is often among the primary missing elements, prompting us to discard this model. Nevertheless, we anticipate that forthcoming surveys with significantly larger sample sizes will mitigate this issue effectively.

The HR10 and HR21 spectra were processed in their normalised states, resulting in spectra containing 17,000 pixels. Consequently, the algorithms used are not particularly fast. As we have not yet reached the Nyquist-Shannon sampling level, there is significant potential to speed up the algorithm by following this criterion, retaining the structure while adjusting the hyperparameters accordingly.

While \texttt{OssicoNN} has demonstrated significant effectiveness in addressing neural network challenges, such as providing uncertainty measures, it still faces limitations inherent to supervised machine learning when applied to observational data. Primarily, \texttt{OssicoNN} relies on the input labels provided by GES, which may not always be accurate. 
Furthermore, when applied to stars that were excluded from the training due to incomplete parameter derivation, the resulting estimates are unreliable. This highlights the a limitation of algorithms trained on observational data: they are inherently limited to estimating quantities within the realm of existing knowledge, rendering them less effective tools for exploration and discovery.

We attempted to evaluate the inference on a different dataset, specifically the AMBRE synthetic spectra \citep{ambre}, which are tailored to match the GIRAFFE HR10-HR21 spectral shape. Regrettably, the results were highly unsatisfactory (e.g. the typical residual deviation is around 1000K for effective temperature). 
Therefore, it is more appropriate to consider \texttt{OssicoNN} as an estimator of the parameters assigned by GES for each star, rather than an independent parameter estimator.

Solving all this issues necessitates the exploration of various paths, including the advancement of domain adaptation techniques to minimise disparities between synthetic datasets (these techniques can leverage synthetic datasets encompassing a wider range of parameters, even those not yet observed) and observed datasets. Another possibility would be the exploration of unsupervised methods, which exhibit greater adaptability across multiple datasets.

\section{Summary and conclusions}
\label{sec_conclusion}
In this paper, we introduce \texttt{OssicoNN}, a neural network expressly devised to address challenges associated with the determination of stellar parameters. At its core lies the application of the conditional invertible neural network (cINN), an elegant framework developed by \cite{Ardizzone_INN}. Leveraging the flux uncertainty present in the spectra and the properties of cINN, \texttt{OssicoNN} offers an estimation of the overall uncertainty derived during the inference process, encompassing both the internal error associated with the neural network and the external error intrinsic to the data. \texttt{OssicoNN} is trained on stellar observational data from the GIRAFFE dataset of the \textit{Gaia-ESO} Survey to  derive effective temperature, surface gravity, metallicity, and various elemental abundances (aluminium, magnesium, calcium, nickel, titanium, and silicon) from stellar spectra.

The appeal of cINNs lies in their ability to compute Bayesian posterior distributions for all parameters $x$ given an observation $y$. This is facilitated by introducing auxiliary variables known as the latent space $z$, which serve as a reservoir of diversity built during training. During inference, this diversity is constrained by the observation to retain only the feasible possibilities. 

The training set comprises 6688 high-quality stars ($S/N>25$). The spectra encompass two spectral domains: HR10 ($5330-5610 \AA$, $R=21500$) and HR21 ($8480-8980 \AA$, $R=18000$). The training set predominantly includes stars with effective temperatures ranging $T_{eff}=4000-6915$K, surface gravities ($\mathrm{log}⁡(g)$) spanning from 0.48 to 4.84 dex, and metallicities ([Fe/H]) varying from $-$1.5 to 0.47 dex.

We validated the performance of \texttt{OssicoNN} in various methods. Our main results can be summarised as follows:

\begin{enumerate}[1.]
    \item The neural network demonstrates exceptional accuracy on the test set (S/N>25), achieving an accuracy of 28K in $T_{eff}$, 0.06 dex in $\log (g)$, 0.03 dex in [Fe/H], and between 0.05 dex and 0.17 dex for other abundances.

    \item Accuracy remains stable even for spectra with low S/N (of between 5 and 25), with an accuracy of 39K in $T_{eff}$, 0.08 dex in $\log (g)$, and 0.05 dex in [Fe/H].

    \item To identify the spectral channels critical for inferring each parameter, we employed a saliency map (gradient analysis). The results indicate that the determination of abundance is based on the absorption lines of the element under consideration, such as the $8892.720 \AA$ line for silicon or the  $8773.896\AA$ and $8773.896\AA$  doublet for aluminium. This is not the case only for titanium, which is primarily determined by lines originating from other elements. 
    
    \item Utilising the capabilities of cINN, we can determine the distribution of parameter posteriors for a spectrum, thereby quantifying internal uncertainties. By employing the Monte Carlo method to vary the spectra, accounting for the variance of each spectrum, the external error is measured. The total error is then computed as the quadratic sum of the individual errors.

    \item The error magnitude is comparable to the accuracy, with a median uncertainty of 55 K for temperature, 0.08 dex for $\log (g)$, 0.05 dex for [Fe/H], and between 0.06 dex and 0.17 dex for chemical abundances.
    
    \item The behaviour of the uncertainties aligns with expectations, which is evident in the residual/uncertainty ratio, which follows a Gaussian distribution $\mathcal{N}(0,1)$, as predicted by the central limit theorem, assuming the uncertainties are independently drawn from a normal distribution and the relationship between GES and \texttt{OssicoNN} estimates is linearly homoscedastic. Additionally, the uncertainties are consistent with those determined by GES.

    \item The study of external and internal uncertainties allowed us to identify the key parameters that contribute to error determination, even indirectly. The external error is strongly related to the S/N. The internal error depends on the distribution of the training set, potential degeneracies, and internal parameters of the neural network (establishing a precise relationship is more straightforward in the central regions of the parameter space than at its edges).

    \item We then verified that the neural network accurately reproduces astrophysical relationships at both the Milky Way scale and within small star clusters. In the Milky Way, this verification was conducted at two levels. First, we examined the chemical separation between the stars of the thin disc and the thick disc in the [Mg/Fe]--[Fe/H] plane. \texttt{OssicoNN} achieves a clearer separation between the two discs compared to the labels in the GES catalogue. Second, we analysed the evolution of stars in the Milky Way using the Kiel diagram. The Kiel diagram of parameters inferred by \texttt{OssicoNN} correctly depicts the various branches and sequences. Additionally, there is an improvement in the Kiel diagram, with some low-S/N stars being correctly repositioned with their appropriate metallicities. However, there are some weaker areas, particularly in regions where the training set is sparsely populated, leading to less precise inferences. This shortcoming is mitigated when the training set is artificially expanded.

    \item Within the clusters, the relationships are accurately reproduced. The anticorrelation observed in globular clusters is well replicated. In open clusters, the parameters exhibit a tight Gaussian dispersion of abundances, with the distribution being 33\% narrower than that of the labels. Additionally, stars plotted on the Kiel diagram follow the isochrone (fixed age--metallicity) of the cluster.

    \item We compared \texttt{OssicoNN} with independent estimates of benchmark stars from \cite{jofre:2015}. The two sets of estimates are similar when the parameters fall within the training set range; otherwise, deviations occur, such as at very low metallicity ([Fe/H]<-1.5). Within the training set range, the precision is -52 K for temperature, 0.04 dex for surface gravity, and less than 0.06 dex for abundance. Expanding the training set with augmentation resolves the issue of an insufficiently broad training range.

    \item A new GES HR10 and HR21 survey catalogue has been derived using \texttt{OssicoNN}.
\end{enumerate}

All these points emphasise that \texttt{OssicoNN} is highly suited for application with new instruments such as 4MOST and WEAVE.

\section{Data availability}
\label{section_data_availability}
The additional figures are available on Zenodo : \href{https://zenodo.org/records/13844177}{https://zenodo.org/records/13844177}.
The code is available on GitHub : \href{https://github.com/nils-can/OssicoNN}{https://github.com/nils-can/OssicoNN}.
The weights of the driven hyperparameters are available on request.
The new GES HR10 and HR21 survey catalogue derived using \texttt{OssicoNN} is available at the CDS. 
\begin{acknowledgements}
      This work was supported by the INAF Techno grant 2022 MALSPEC (MAchine Learning for SPECtroscopy, PI: Sacco). This work is based on data obtained from the ESO Science Archive Facility with DOI(s): https://doi.org/10.18727/archive/25 (The \textit{Gaia-ESO} Survey, co-PI: G. Gilmore and S. Randich).

      We would like to express our sincere gratitude to the anonymous reviewer for their thoughtful and constructive comments. Their insightful feedback has significantly contributed to the improvement of this manuscript.

\end{acknowledgements}

%
%

\bibliographystyle{aa}
\bibliography{biblio}

\begin{appendix} 
\onecolumn
\section{Additional Figures}
    \begin{figure*}[!htb]
   \centering
   \includegraphics[width=\hsize]{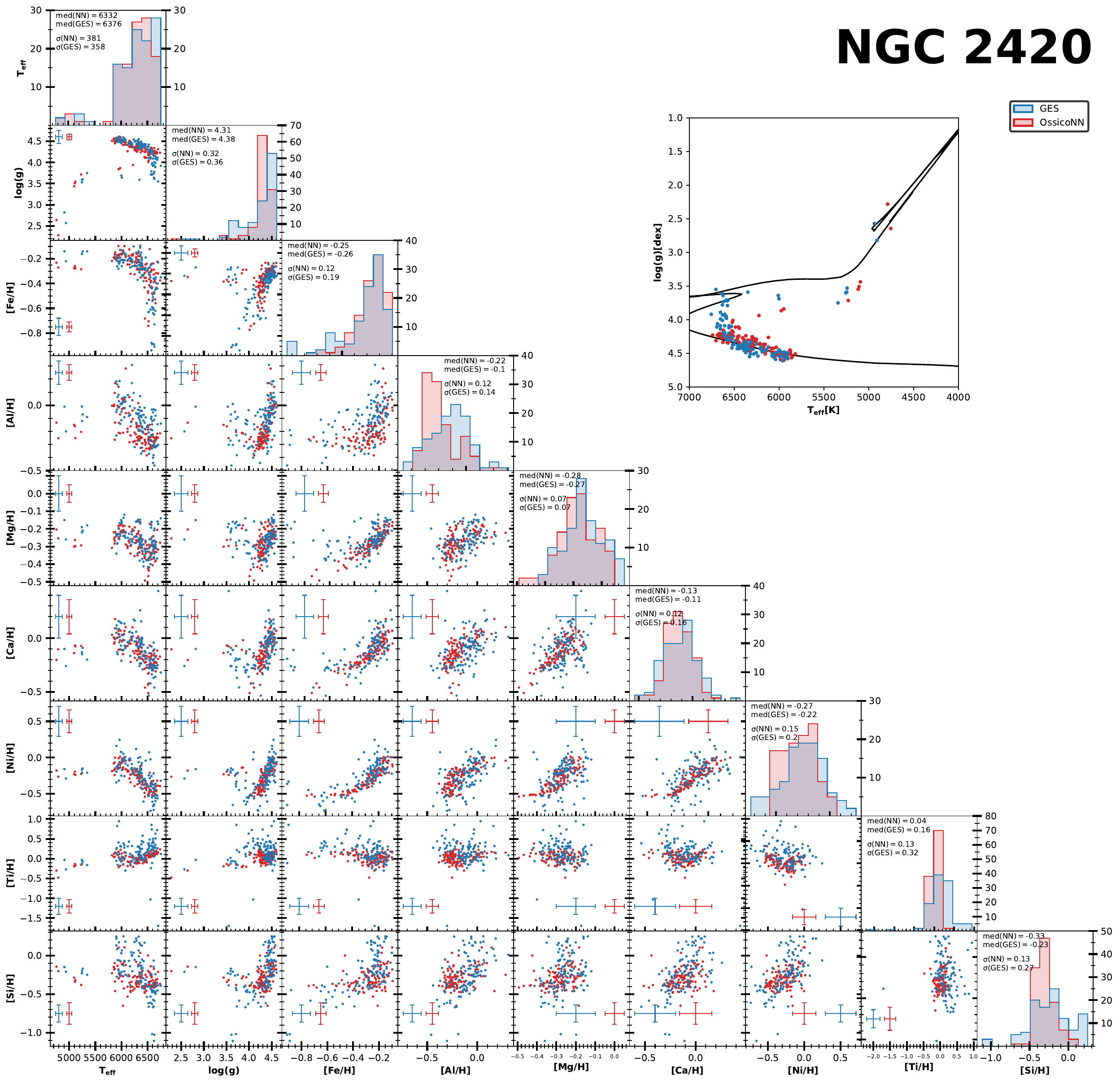}
      \caption{Parameters and abundances of stars belonging to the NGC2420 cluster using the GES pipeline and the \texttt{OssicoNN} neural network. The number of GES stars varies across quadrants since it depends on how many stars could be successfully processed through the classical pipelines for the given set of parameters. We also display the Kiel diagram and the isochrone corresponding to the age and metallicity of the cluster measured by \cite{cantat-gaudin:2020}. The isochrone is generated using PARSEC version 1.2S (\cite{parsec_bressan})., \cite{parsec_chen}).}
         \label{fig:open_cluster_NGC2420}
   \end{figure*}

   \begin{figure*}[]
   \centering
   \includegraphics[width=\hsize]{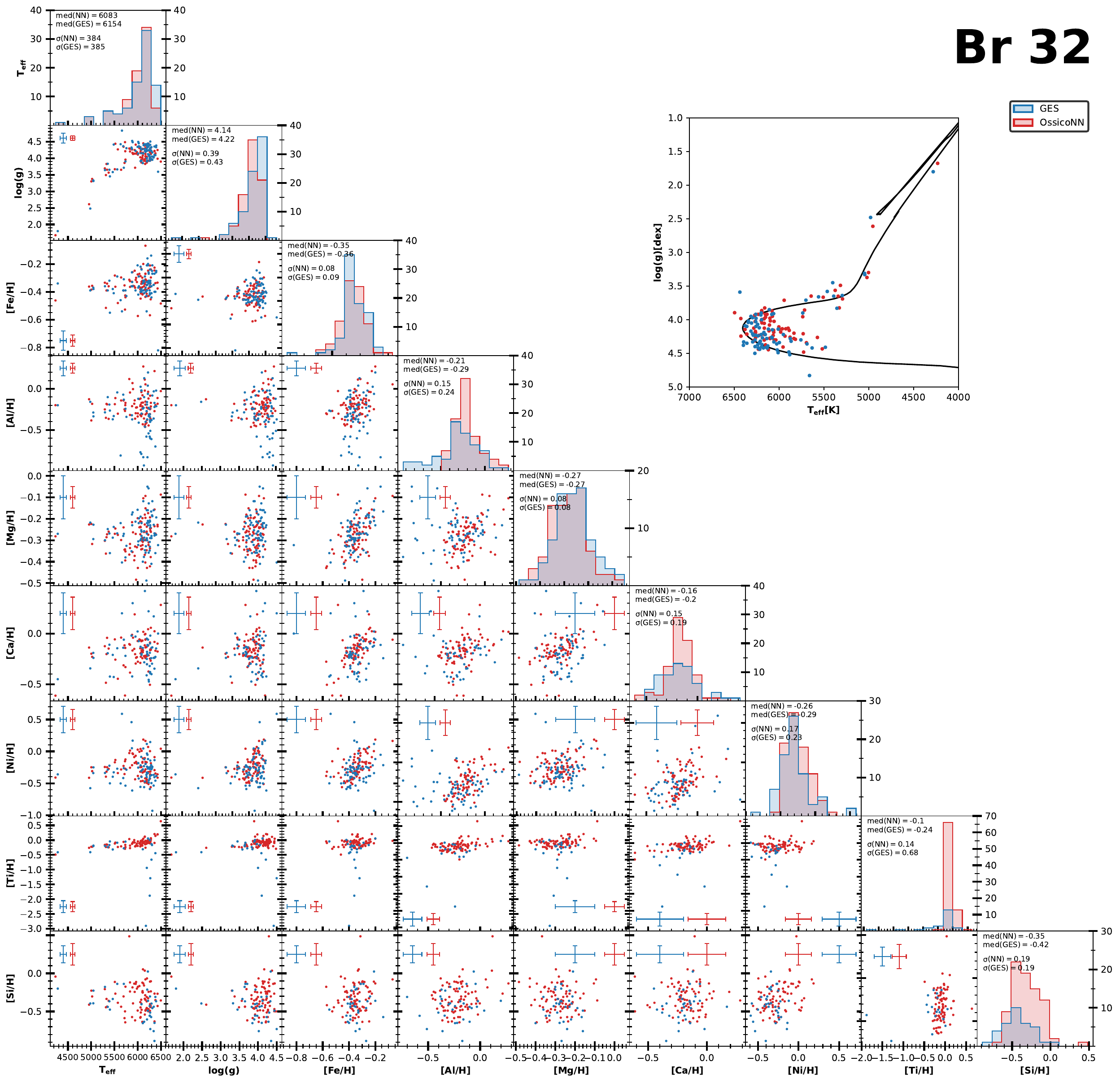}
      \caption{Parameters and abundances of stars belonging to the Br32 cluster using the GES pipeline and the \texttt{OssicoNN} neural network. The number of GES stars varies across quadrants since it depends on how many stars could be successfully processed through the classical pipelines for the given set of parameters. We also display the Kiel diagram  and the isochrone corresponding to the age and metallicity of the cluster measured by \cite{cantat-gaudin:2020}. THe isochrone is generated using PARSEC version 1.2S (\cite{parsec_bressan}, \cite{parsec_chen}).}
         \label{fig:open_cluster_br32}
   \end{figure*}
\clearpage
\section{Latent space study}
\label{sec:latent_space_study}

This section contains figures relating to the study of latent space and internal errors. First, we focus on how the posterior distribution of the parameters (obtained by sampling the latent space) are related in pairs for three stars. We measured the Pearson coefficients for the stars in the test set and obtained the following distributions, which are very little correlated.
\begin{figure*}[h]
    \resizebox{\linewidth}{!}
              {\includegraphics{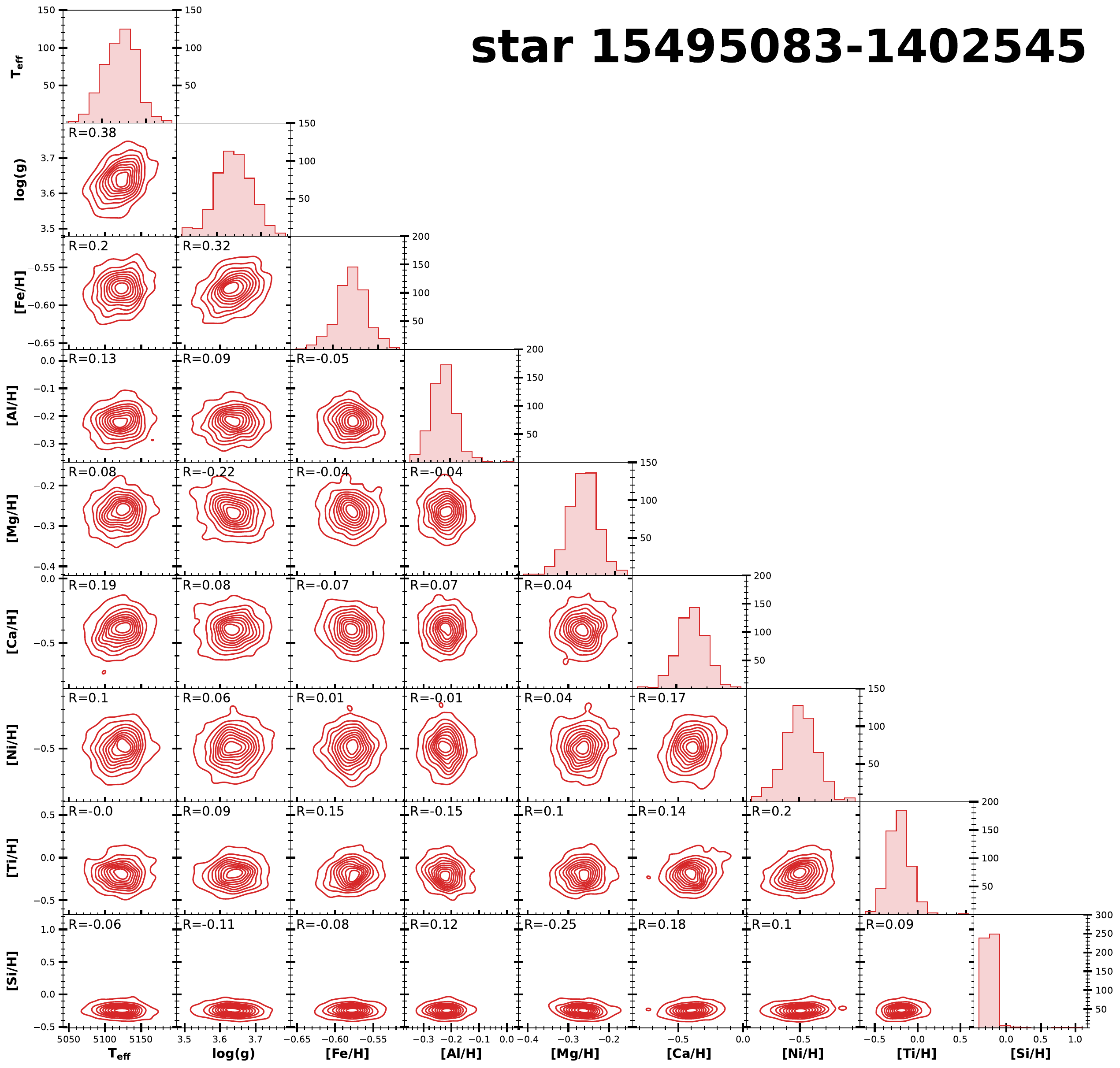}}
    \caption{Pairwise distribution of all parameters from Latent Space sampling using \texttt{OssicoNN}  for star 15495083-1402545 using $N_{\mathrm{internal}}=500$ samples. Pearson coefficient is computed for each distribution.}
    \label{fig:distribution_latent_space_20}
\end{figure*}

\begin{figure*}
    \resizebox{\linewidth}{!}
              {\includegraphics{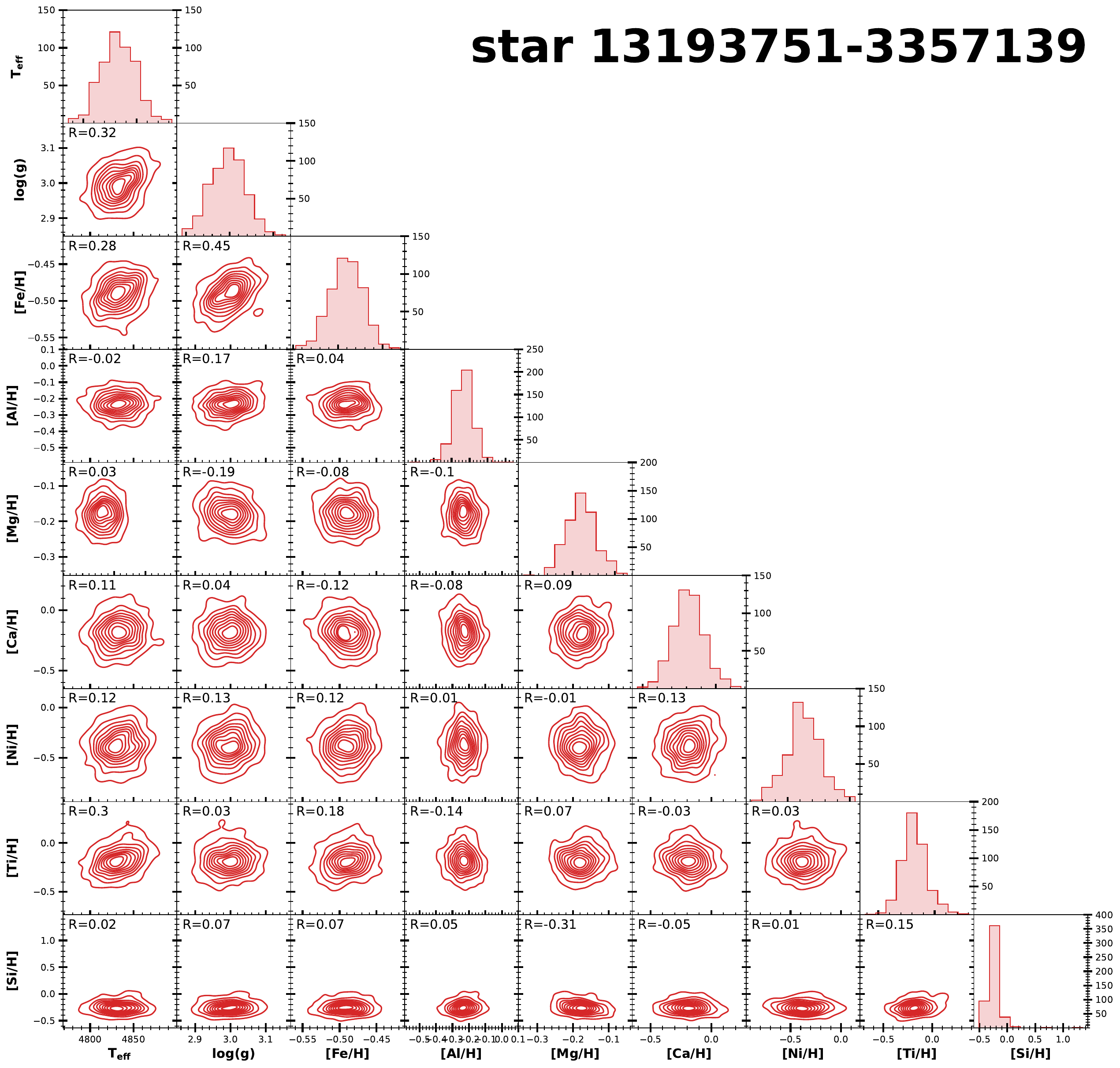}}
    \caption{Pairwise Distribution of all parameters from Latent Space sampling using \texttt{OssicoNN}  for star 13193751-3357139 using $N_{\mathrm{internal}}=500$ samples. Pearson coefficient is computed for each distribution.}
    \label{fig:distribution_latent_space_250}
\end{figure*}

\begin{figure*}
    \resizebox{\linewidth}{!}
              {\includegraphics{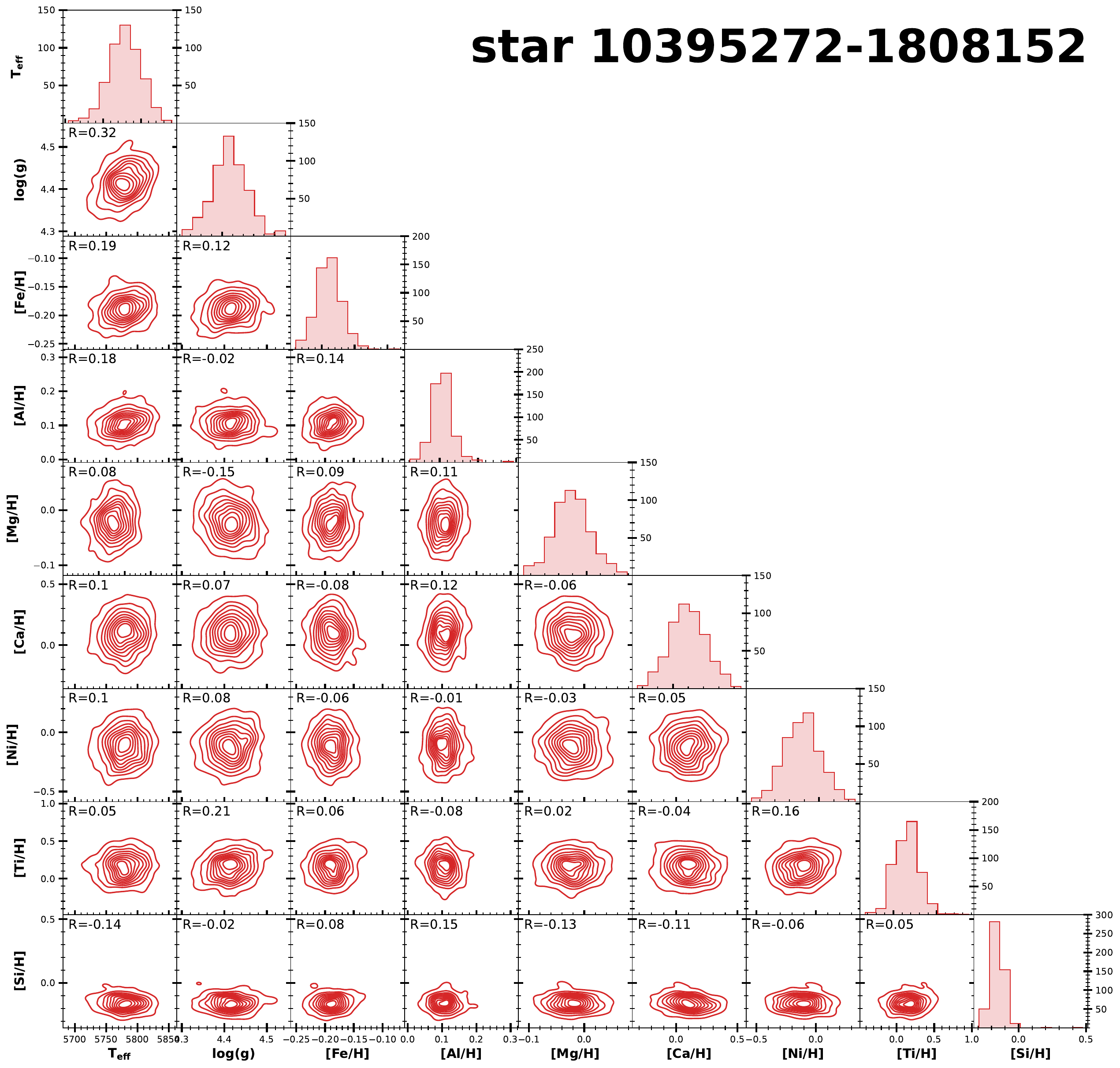}}
    \caption{Pairwise Distribution of all parameters from Latent Space sampling using \texttt{OssicoNN}  for star 10395272-1808152 using $N_{\mathrm{internal}}=500$ samples. Pearson coefficient is computed for each distribution.}
    \label{fig:distribution_latent_space_750}
\end{figure*}

\begin{figure*}
    \resizebox{\linewidth}{!}
              {\includegraphics{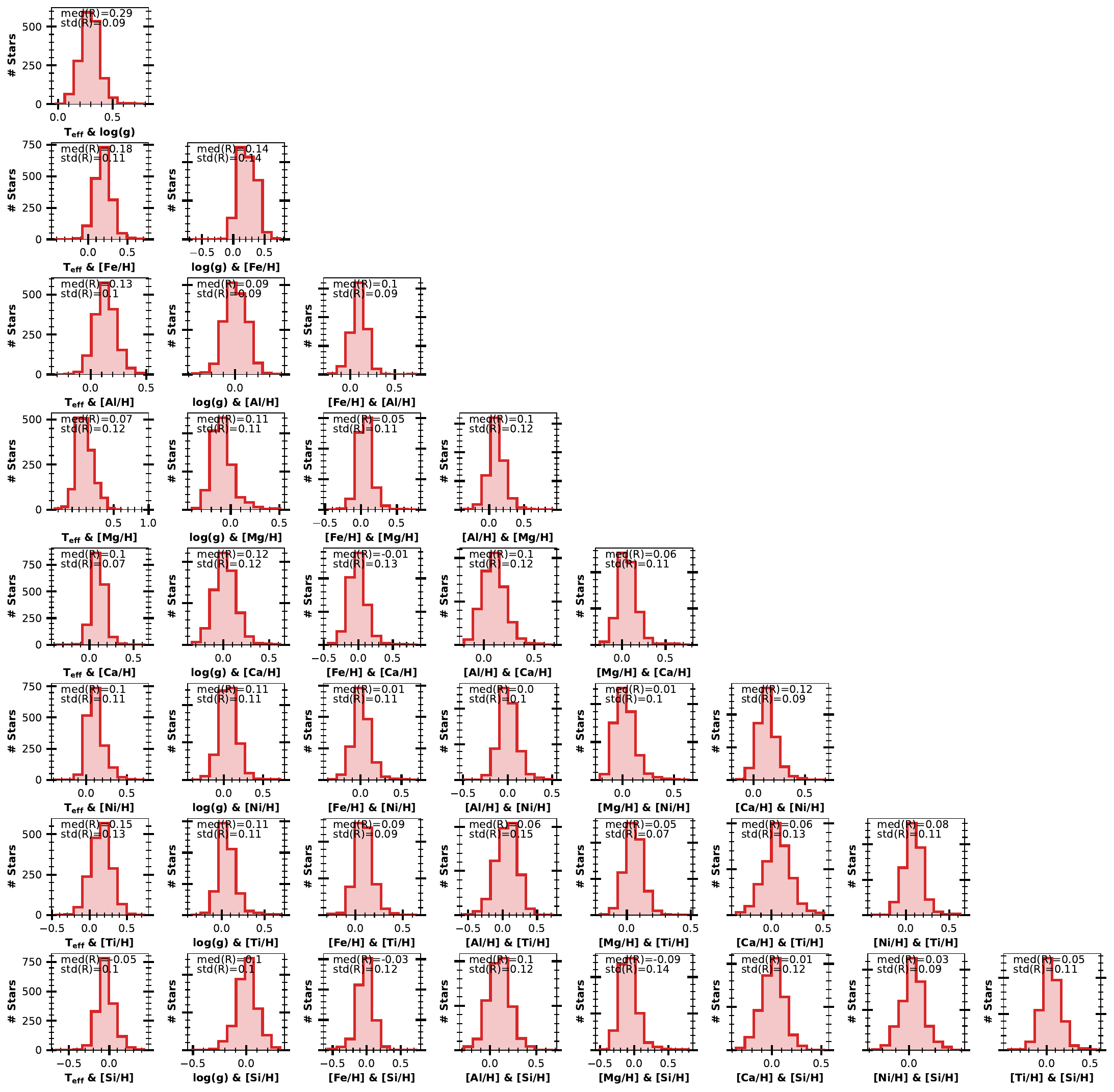}}
    \caption{Distribution of Pearson coefficients between pairs of parameters for test set stars, derived from sampling latent space to obtain the posterior distribution for one spectrum.}
    \label{fig:distribution_pearson_coeff}
\end{figure*}
\clearpage
\section{Uncertainty: additional figures}
\begin{figure*}[h]
    \resizebox{\linewidth}{!}
              {\includegraphics{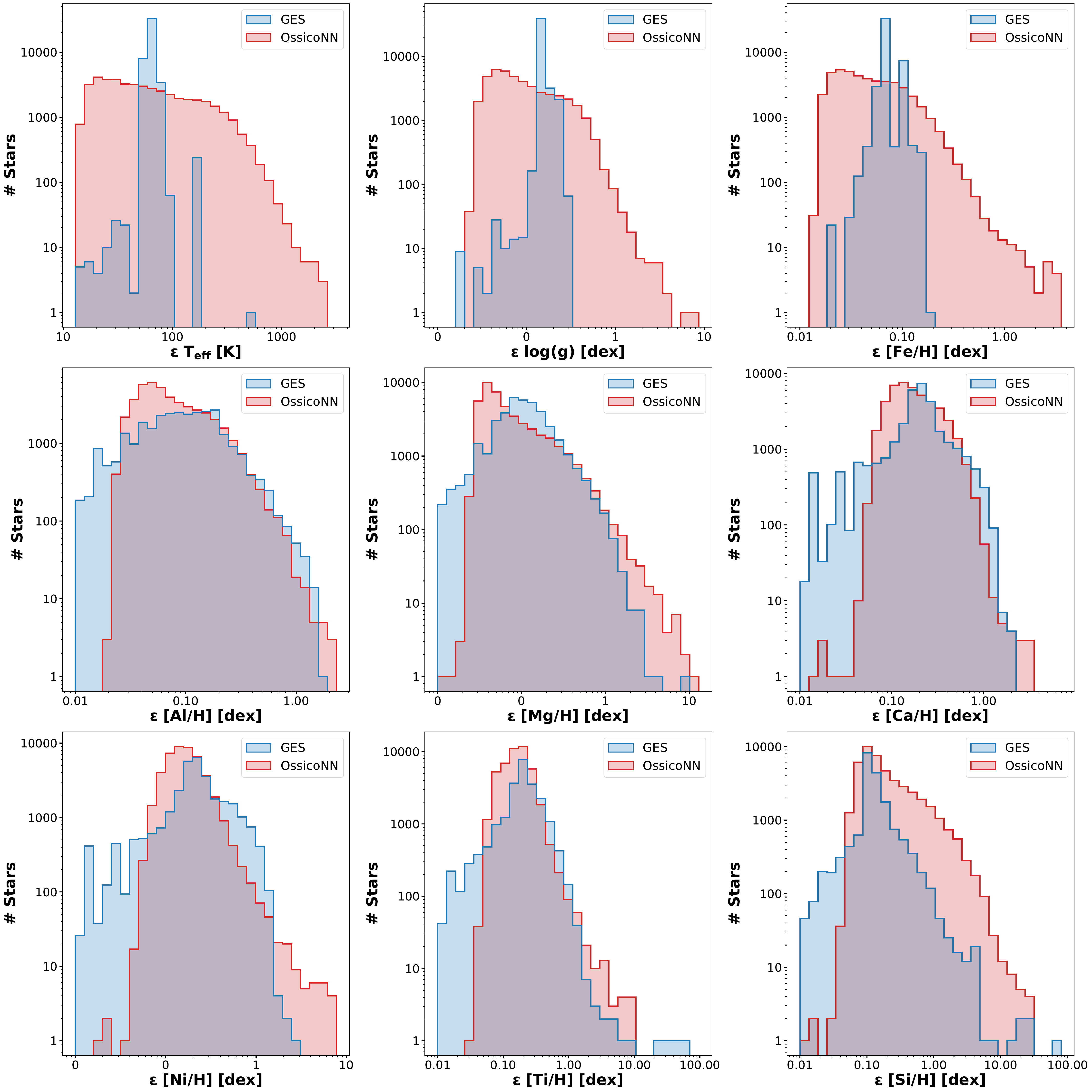}}
   \caption{Distribution of uncertainties for GES(in blue) and OssicoNN (in red) for the Reduced Catalogue dataset (discuss in section \ref{sec_uncertainty}.)}
   \label{fig:distribution_error}
\end{figure*}
    \begin{figure*}
    \resizebox{\linewidth}{!}
              {\includegraphics{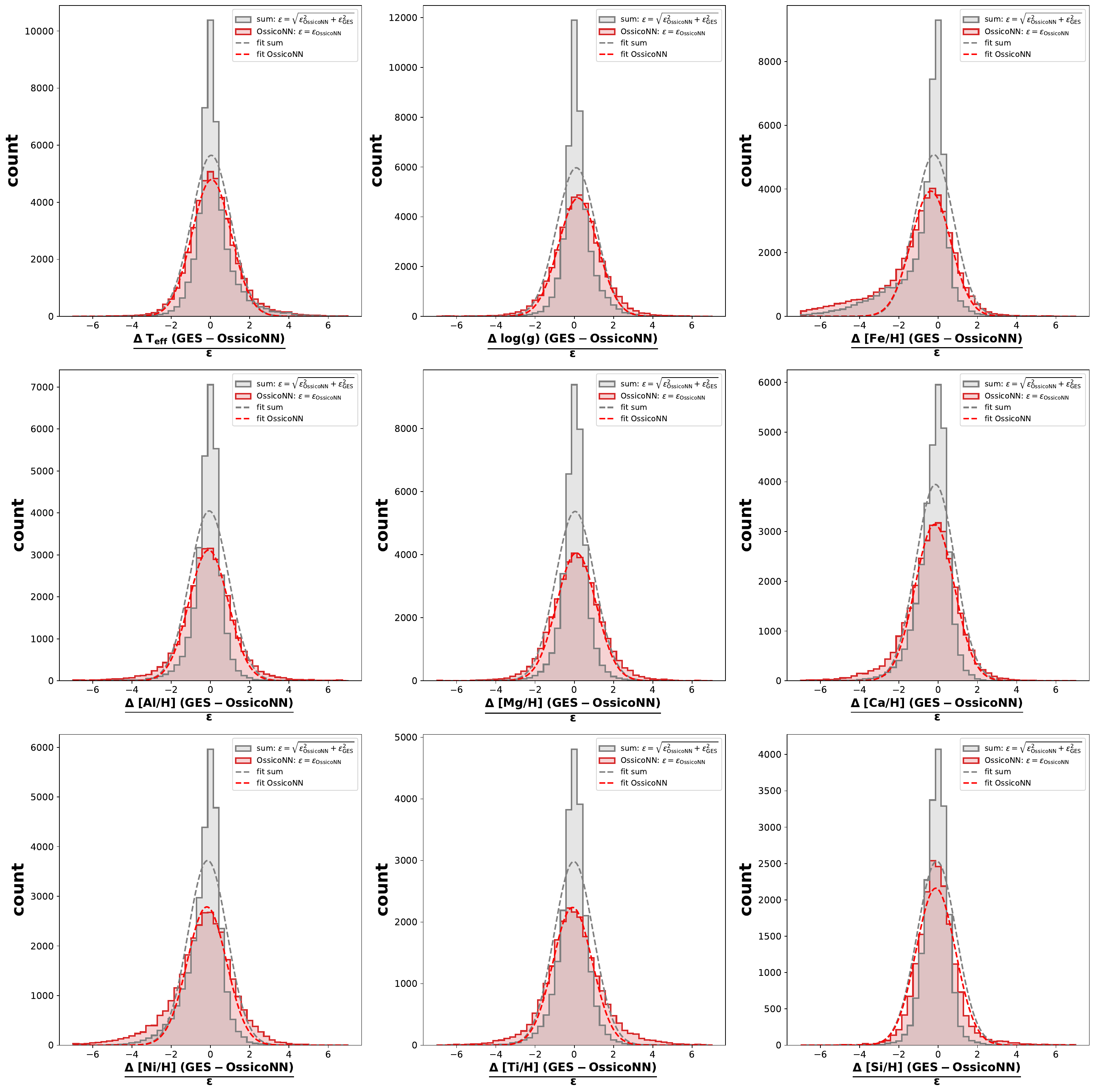}}
    \caption{Distribution of the residuals between \texttt{OssicoNN} and GES, normalised with respect to the uncertainties of \texttt{OssicoNN} (red-shaded area) and the sum in quadrature of GES and \texttt{OssicoNN} errors (grey-shaded area) for the Reduced Catalogue dataset. The dashed lines represent the fit of the distribution using a Gaussian function with a standard deviation of $\sigma = 1$.}
    \label{fig:distribution_error_gauss}
    \end{figure*}

    \begin{figure*}
    \resizebox{\linewidth}{!}
              {\includegraphics{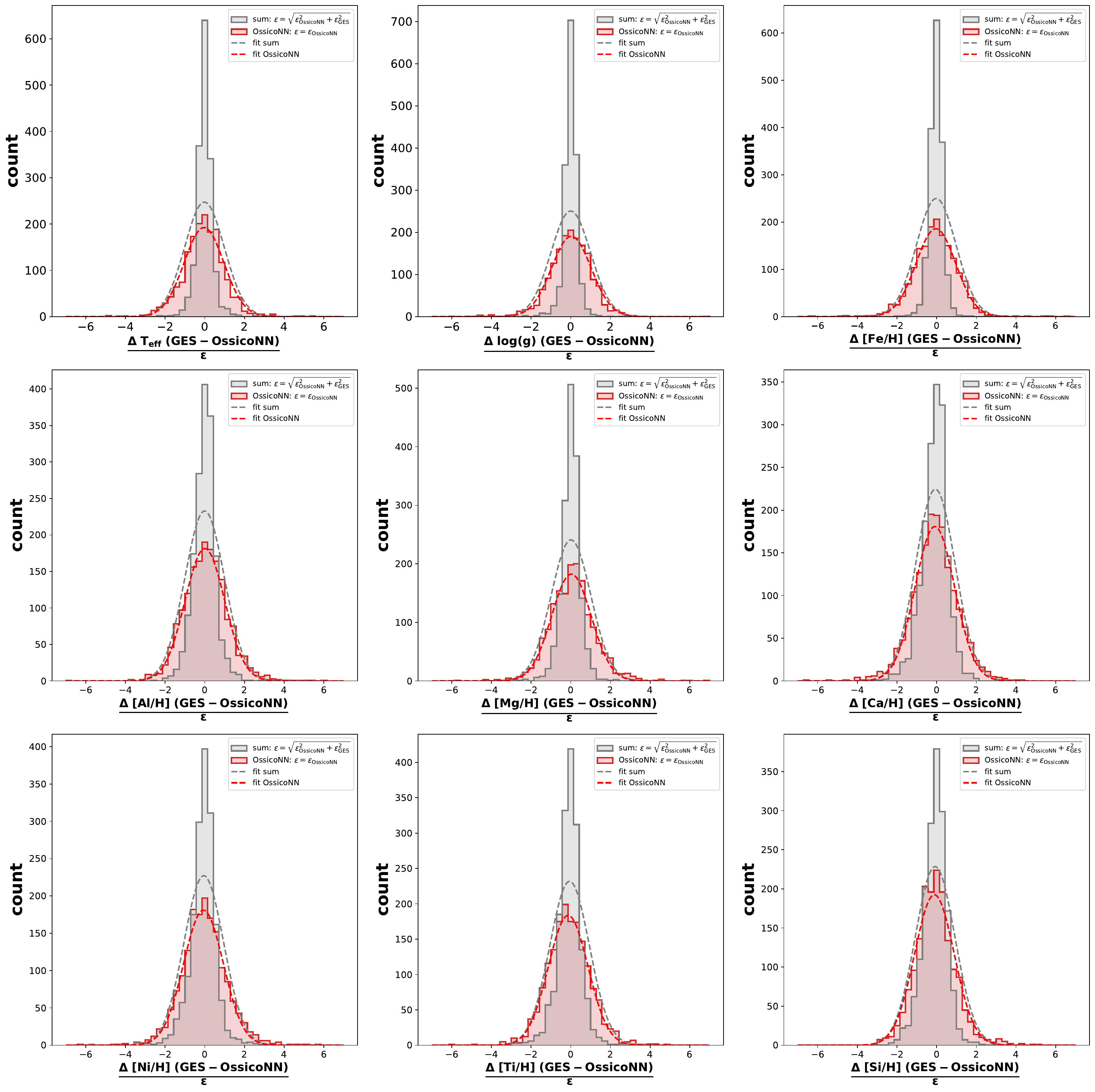}}
    \caption{Distribution of the residuals between \texttt{OssicoNN} and GES, normalised with respect to the uncertainties of \texttt{OssicoNN} (red-shaded area) and the sum in quadrature of GES and \texttt{OssicoNN} errors (grey-shaded area) for the Test set. The dashed lines represent the fit of the distribution using a Gaussian function with a standard deviation of $\sigma = 1$.}
    \label{fig:distribution_error_gauss_test}
    \end{figure*}

    \begin{figure*}
    \resizebox{\linewidth}{!}
              {\includegraphics{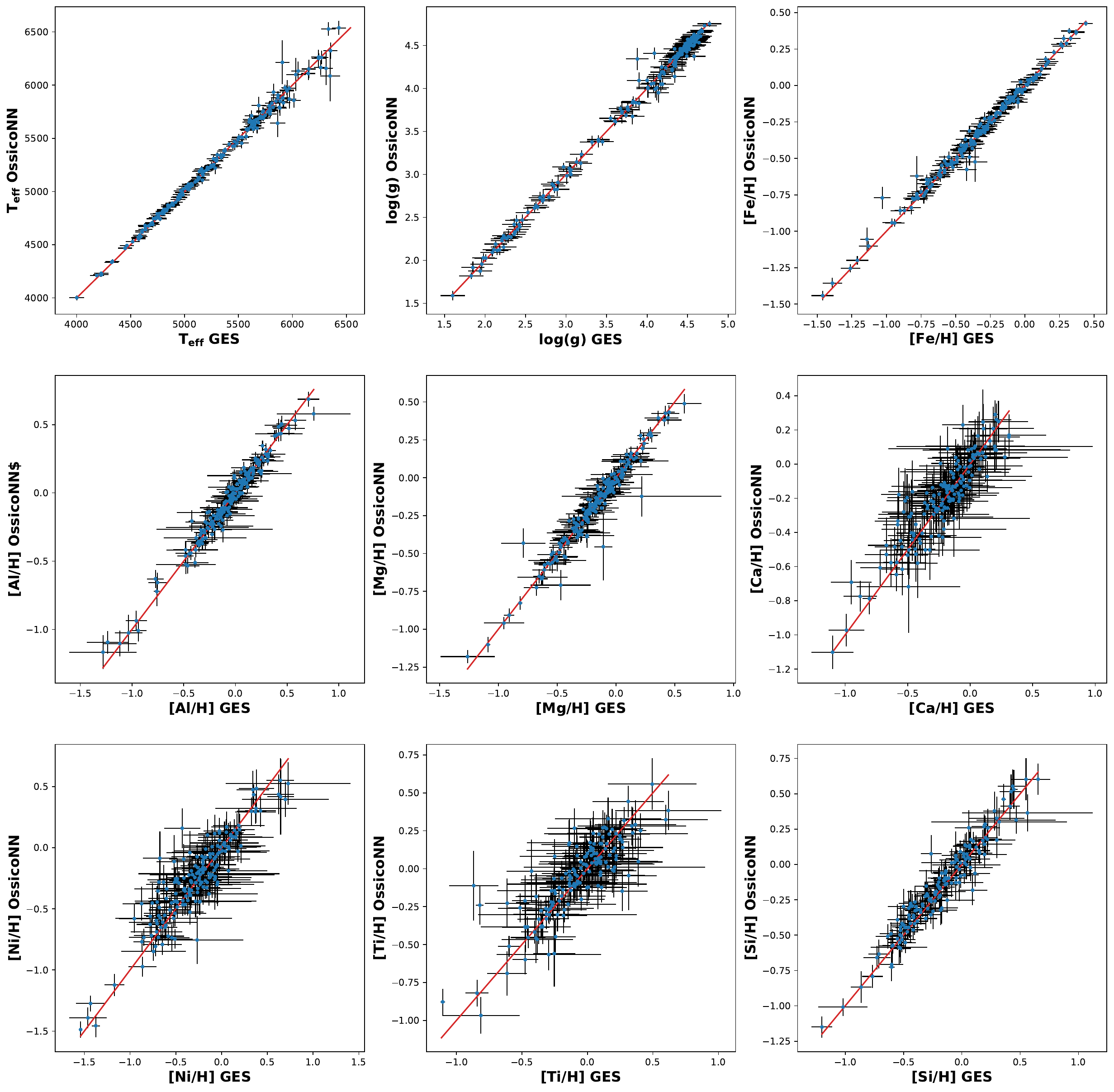}}
    \caption{Parameter estimates for 200 randomly selected stars from the Reduced Catalogue dataset (with restrictions discussed in Section \ref{sec_uncertainty}) by GES and \texttt{OssicoNN}, along with their respective uncertainties.}
    \label{fig:uncertainties_one_one}
    \end{figure*}

    \begin{figure*}
    \resizebox{\linewidth}{!}
              {\includegraphics{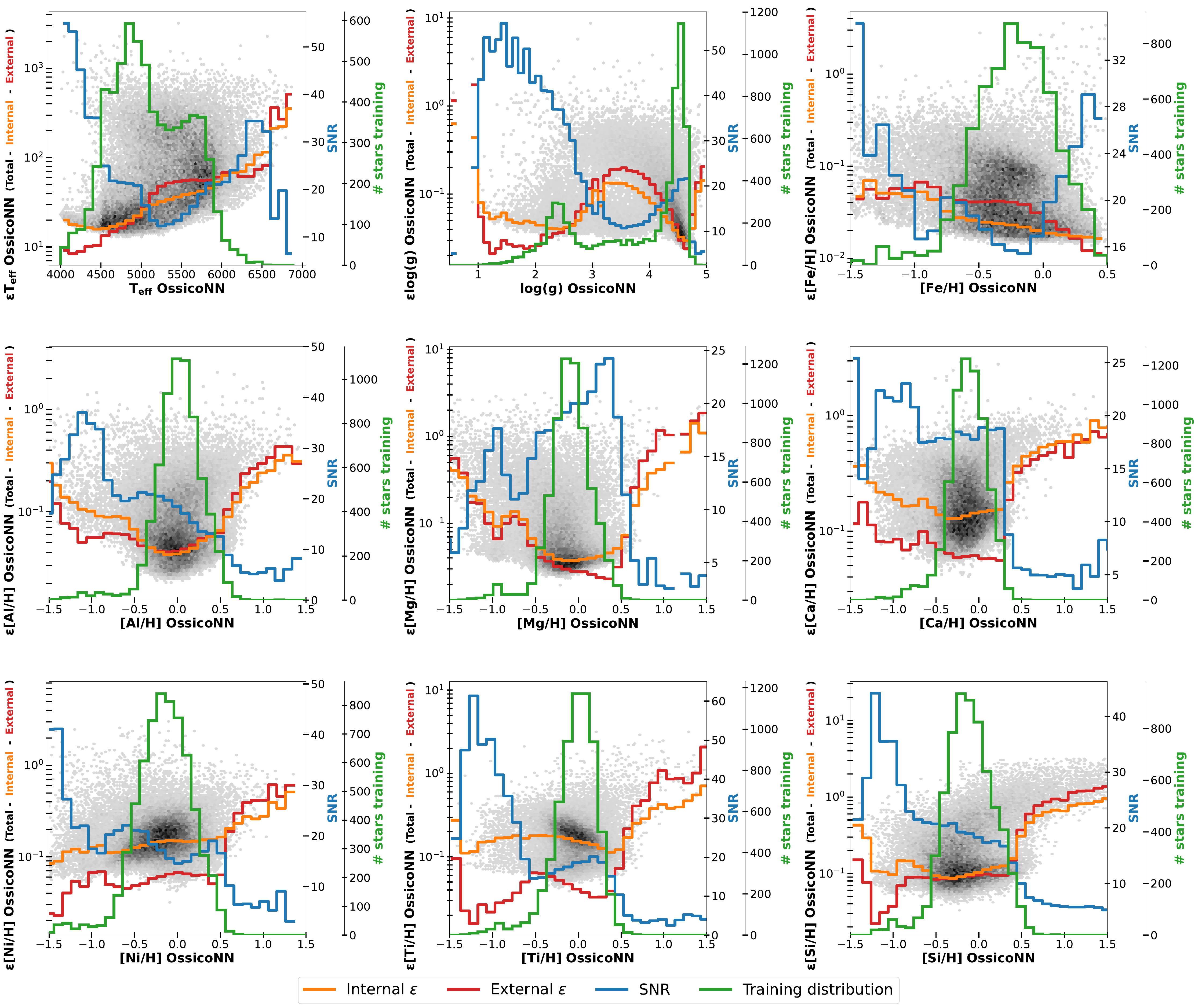}}
    \caption{Density distribution of total \texttt{OssicoNN} uncertainties relative to the inferred parameters for the Reduced Catalogue dataset. The red and orange lines indicate the median external and internal uncertainties across parameter bins. The blue and green distributions, with scales on the right, represent the median S/N and the training set distribution per parameter bin, respectively.}
    \label{fig:uncertainties_alea_epistemic}
    \end{figure*}

\clearpage
\section{Astrophysical relation: additional figures}

    \begin{figure*}[h]
    \centering
    \resizebox{0.87\linewidth}{!}
              {\includegraphics{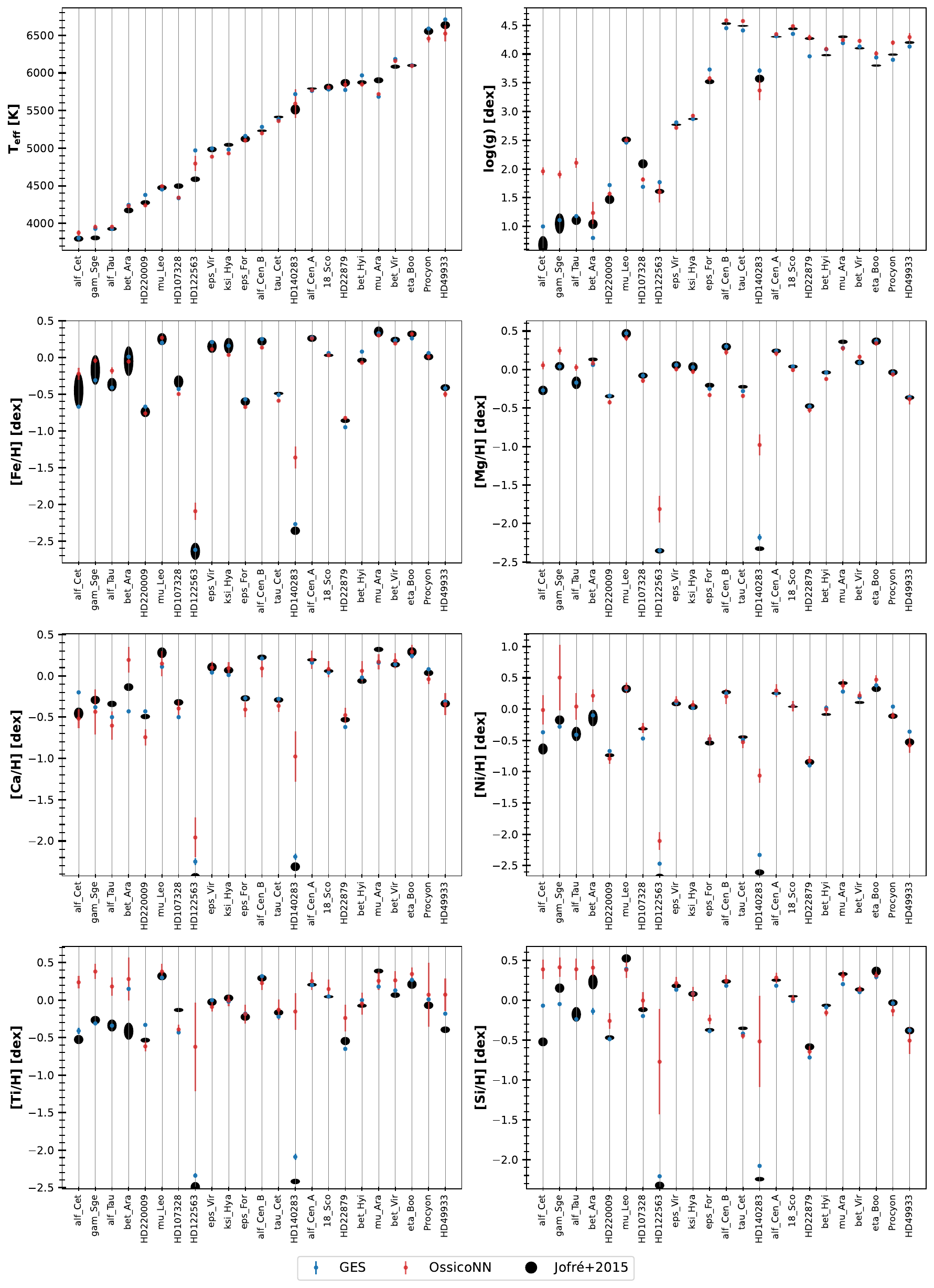}}
    \caption{Parameter estimation for selected Benchmark stars using the GES pipeline, the \texttt{OssicoNN} neural network and \cite{jofre:2015}, which combines spectral and spectral-independent analysis. The vertical half-axes of the Jofré+2015 points indicate the uncertainty of the estimate.}
    \label{fig:benchmark_all}
\end{figure*}
\twocolumn   
\section{Maximum likelihood loss}
\cite{Ardizonne:cINN} suggest training cINNs by formulating a loss function that considers both the input and latent space parameters, along with the model's  parameters ($\theta$). This loss is constructed based on the likelihood function. Herein, we provide the problem definition and the calculations necessary to derive equation \ref{loss_text}.

Let $p_Z(\mathcal{Z})$ and $p_X(\mathcal{X})$ be probability densities defined on the spaces of $\mathcal{Z}$ and $\mathcal{X}$, respectively. Let $\theta$ denote the parameters of a neural network, and let $y$ be a conditioning variable. Then, the neural network function is given by $f(x|\theta, y )$, which maps an input $x$ to an output depending on $\theta$ and $y$.

Given a set of training data $(x_{i})_{i=1}^{n}\in \mathcal{X}^{n}$, we can formulate the likelihood function as:
\begin{equation}
L(\theta) \doteq \prod_{i=1}^{n}p_{\theta}(x_i).  
\end{equation}

The likelihood function measures how well the model parameters $\theta$ fit the observed data. A common approach to estimate $\theta$ is to maximise the likelihood function. The change-of-variable formula enables us to establish a connection between the probabilities in the data space and the probabilities in the latent space:
\begin{align}
p_X(\mathbf{x};\mathbf{c},\theta) &= p_Z(\mathbf{z};\mathbf{c},\theta) \left|\det\left(\frac{\partial f}{\partial x}\right)\right| \\
    &= p_Z(f(\mathbf{x};\mathbf{c},\theta)) \left|\det\left(\frac{\partial f}{\partial x}\right)\right|
\end{align}

Bayes' theorem can be used to determine the posterior along the model parameters:
\begin{equation}
    p(\theta; \mathbf{x, \mathbf{y}}) \propto p_X(\mathbf{x}, \mathbf{y}; \theta) ~p_{\theta}(\theta).
\end{equation}

Starting from the likelihood minimisation and applying the formulas above we obtain :
\begin{equation}
\begin{split}
    \mathcal{L} &= -\log(L(\theta)) \\
    \mathcal{L} &= -\log \left(\prod_{i=1}^{n}p(\theta;x_{i})\right) \\
    \mathcal{L} &= \mathbb{E}_{i}[-\log(p(\theta;x_{i}))] \\
    \mathcal{L} &= \mathbb{E}_{i}[-\log(p_X(x_i;\theta,y_i)p_\theta(\theta))] \\
    \mathcal{L} &= \mathbb{E}_{i}[-\log(p_X(x_i;\theta,y_i) - \log(p_\theta(\theta))] \\
    \mathcal{L} &= \mathbb{E}_{i}\left[-\log(p_Z(f(x_i;\theta, y_i)) - \log \left( \left| \det\left(\frac{\partial f}{\partial x}\right)\right|\right) \right] - \log(p_\theta(\theta)). 
    \label{loss}
\end{split}
\end{equation}
Assuming that $Z$ and $\theta$ follow a normal distribution
\begin{equation}
\begin{split}
    p_Z(z) & = \exp(-z^2/2)\\
    p_{\theta} &= \exp(-\theta^2/2\sigma^2),
\end{split}
\end{equation}
the result is:
\begin{equation}
    \mathcal{L} = \mathbb{E}_{i}\left[\frac{||f(x_i;\theta, y_i)||^{2}}{2} - \log \left( \left| \det\left(\frac{\partial f}{\partial x}\right)\right|\right) \right] + \frac{1}{2\sigma^2}||\theta||^{2}.
\end{equation}

\section{Augmented dataset \label{Sec:Agu}}
One of the challenges of our machine learning approach is the limited size of the training set. To address this issue, we apply a data augmentation technique that recovers some of the spectra with incomplete parameters. This enables us to include stars that have only a few missing parameters in our training set. However, we also need to ensure that the quality and reliability of the training set are not compromised by adding noisy or inaccurate data. Therefore, we impose some selection criteria based on the importance and availability of the parameters. Temperature, gravity and metallicity are essential for stellar parameter estimation using the classical method, so we exclude any spectra that lack one of these parameters. Likewise, we discard any stars that have more than three missing values out of the six remaining parameters, to avoid introducing too many artificial values. Lastly, we apply a median S/N threshold of 25 to match the quality of the original training set. By applying these criteria, we are able to recover 7,038 spectra, which doubles the size of our training set to 13,726 spectra. We note that these augmented spectra are only used for training purposes and not for testing or validation.

\begin{figure}
    \centering
    \includegraphics[width=\hsize]{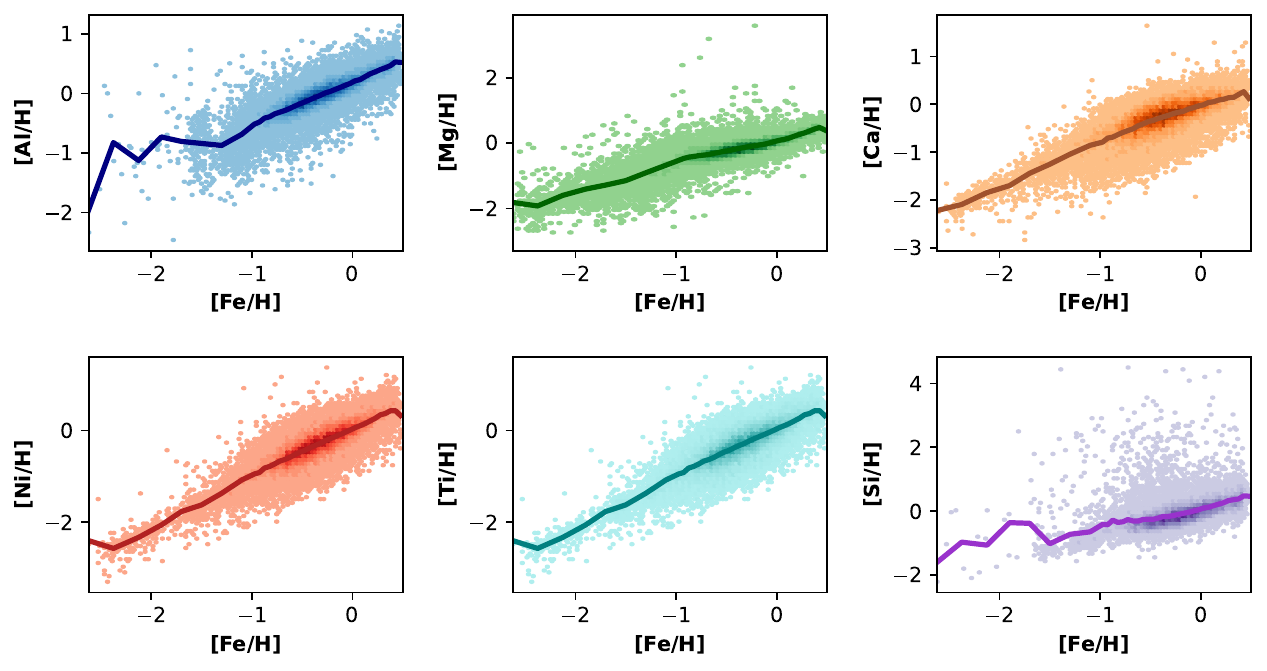}
    \caption{Fit of the element distribution with respect to metallicity, using parameters inferred from classical pipelines and the GES 5.1 data. To fit this distribution, we calculate the average per bin across 39 bins of varying lengths. Specifically, there are 3 bins spanning a range of 0.25 dex between -2.75 and -2 dex, followed by 5 bins between -2 and -1 dex, and finally 31 bins, each with a width of 0.05 dex, covering the range from -1 to 0.6 dex. The average (fit) is represented by the solid line.}
    \label{Fig_fit_element}
\end{figure}

To estimate the missing parameters, we adopt a simple interpolation method based on the metallicity distribution of our data set. We divide the 67,046 spectra into  39 bins according to their metallicity [Fe/H] values, and compute the average element abundance for each bin (Fig. \ref{Fig_fit_element}). Then, we assign this average value to any missing abundance in a spectrum that belongs to that bin. This way, we preserve the correlation between metallicity and element abundance in our data set. 

\begin{figure}
    \includegraphics[width=\hsize]{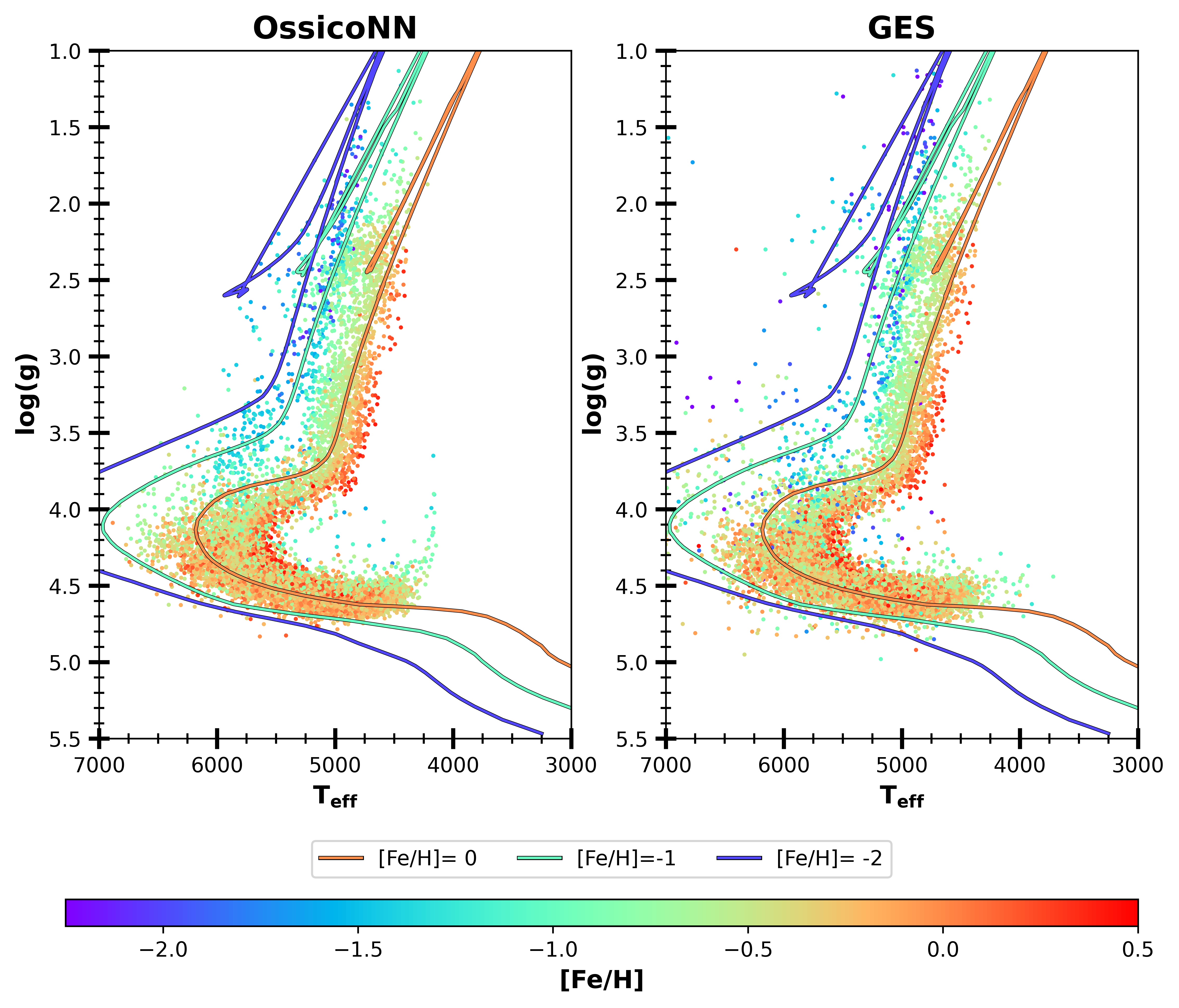}
    \caption{Kiel diagram: log~g versus effective temperature, colour-coded by metallicity, for stars in the Milky Way field with S/N$>$25, as derived using \texttt{OssicoNN} astrophysical parameters (left panel) and the GES recommended values (right panel). Superimposed are three isochrones for an age of 5 Gyr and three metallicities ([Fe/H]=0: orange, [Fe/H=1]: green, [Fe/H=1]: blue) with colours that match the colormap). The isochrones are generated using PARSEC version 1.2S (\cite{parsec_bressan}, \cite{parsec_chen}).}
    \label{fig:kiel_diagram_high}
\end{figure}

\begin{figure}
    \includegraphics[width=\hsize]{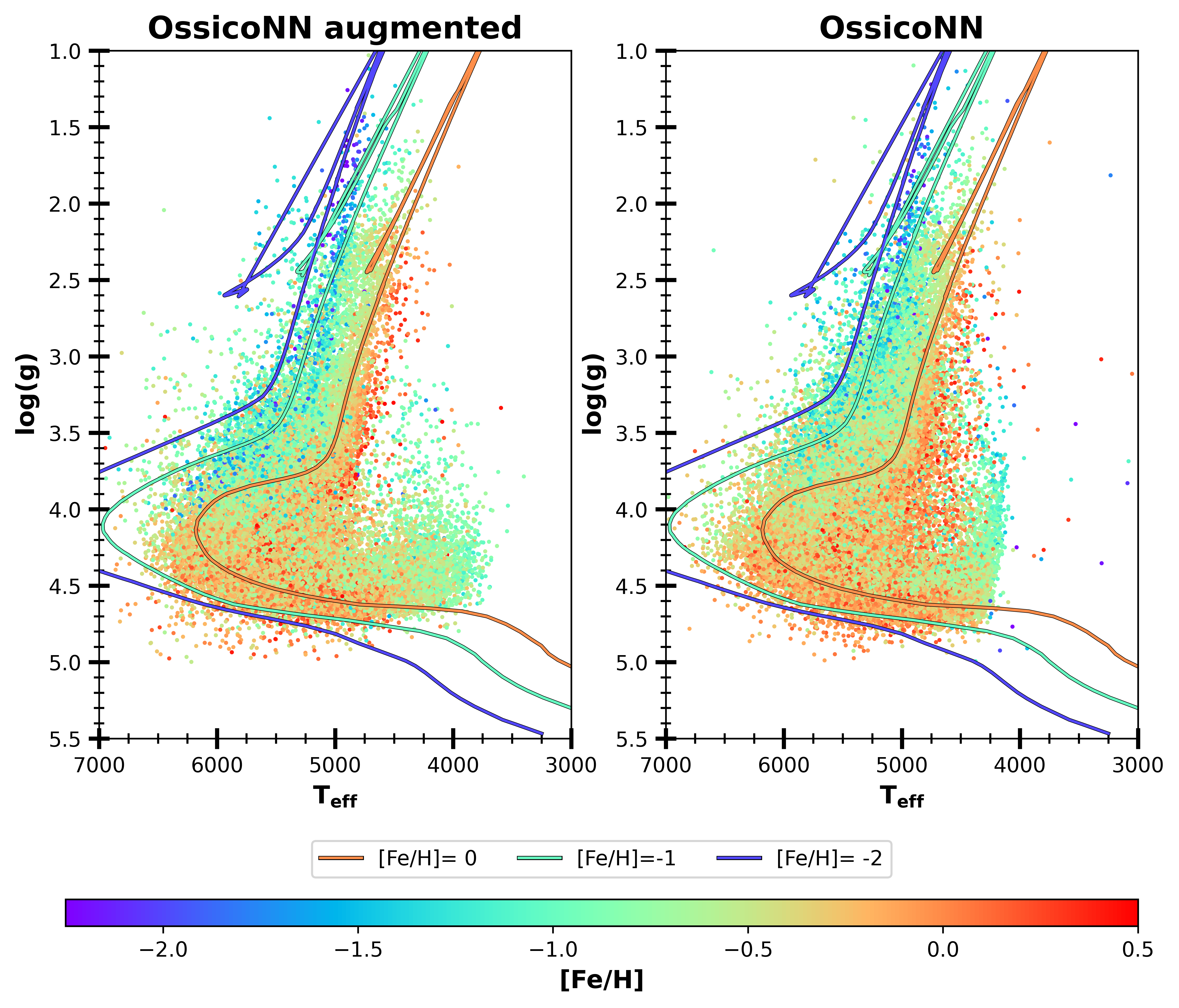}
    \caption{Kiel diagram: log~g versus effective temperature, colour-coded by metallicity, for stars in the Milky Way field as derived using \texttt{OssicoNN} astrophysical parameters with the augmented dataset (left panel) and \texttt{OssicoNN} with the normal training set (right panel). Superimposed are three isochrones for an age of 5 Gyr and three metallicities ([Fe/H]=0: orange, [Fe/H=1]: green, [Fe/H=1]: blue) with colours that match the colormap). The isochrones are generated using PARSEC version 1.2S (\cite{parsec_bressan}, \cite{parsec_chen}).}
    \label{fig:kiel_diagram_augmented}
\end{figure}

Using the augmented dataset not only increases the density of the training data but also extends the range of the training parameters. The temperature range now spans from 3676 to 7205 K, surface gravity ranges from 0.48 to 4.96 dex, and metallicity varies from -2.62 to 0.47 dex. Similarly, the range for aluminium extends from -2.34 to 1.06 dex, magnesium from -2.40 to 0.85 dex, calcium from -2.64 to 1.27 dex, nickel from -2.56 to 1.19 dex, titanium from -2.37 to 2.14 dex, and silicon from -2.21 to 4.39 dex. Augmenting the training set does not impact the precision of the test set, the noise in the dataset, or the uncertainty metrics. The primary changes are observed in the HR10 \& HR21 full dataset, particularly in the Kiel diagram. Previously, the Kiel diagram was highly accurate for high S/N stars (see Fig. \ref{fig:kiel_diagram_high}), but when considering all stars (Fig. \ref{fig:kiel_diag}), certain populations were not well reproduced, such as the end of the red giant branch and the beginning of the main sequence. Additionally, the gradient of the red giant branch was poorly defined for sub-solar metallicities. Fig. \ref{fig:kiel_diagram_augmented} demonstrates that the augmented dataset resolves these issues, with all star populations, including those at the edges, now well defined.

The same phenomenon is observed for the benchmark stars, where all the parameters agree with \cite{jofre:2015}'s estimates. This is particularly evident for HD122563 and HD140283, which have very low metallicities ($-$2.64 and $-$2.36, respectively) and were previously estimated inaccurately as $>-$1.5. With the augmented dataset \texttt{OssicoNN\_agm}, the new estimates for HD122563 and HD140283 are $-$2.64 and $-$2.27, respectively. However, we did not retain this model because, by measuring the correlation coefficients between elements, we found that the augmentation led to deviations from GES by approximately 0.10 for silicon and titanium, the two elements for which we interpolated the most values. Nevertheless, this comparison between augmented and non-augmented datasets underscores that the main issues identified by \texttt{OssicoNN} are due to the training set being too small. These problems are expected to be resolved in new surveys, where both the quantity and quality of data will be significantly higher.

\end{appendix}
\end{document}